\documentclass[fleqn,useAMS]{Papers}
\usepackage{newtxtext,newtxmath}
\usepackage[T1]{fontenc}
\usepackage{subfloat}
\usepackage{makecell}

\usepackage[square,sort,numbers]{natbib}

\usepackage{graphicx}	
\usepackage{amsmath}	
\usepackage{lipsum}
\usepackage{caption}
\usepackage{textgreek} 
\usepackage{xcolor}  
\newcommand{\rev}[1]{{\color{black} #1}} 

\title[Universal scaling laws and halo density slope]{Universal scaling laws and density slopes for dark matter haloes}

\author[Z. Xu]{Zhijie (Jay) Xu,$^{1}$\thanks{E-mail: \href{mailto:zhijie.xu@pnnl.gov}{zhijie.xu@pnnl.gov}; \href{mailto:zhijiexu@hotmail.com}{zhijiexu@hotmail.com}}
\\
$^{1}$Physical and Computational Sciences Directorate, Pacific Northwest National Laboratory; Richland, WA 99354, USA\\
}

\date{Accepted XXX. Received YYY; in original form ZZZ}

\pubyear{2022}

\begin{document}
\label{firstpage}
\pagerange{\pageref{firstpage}--\pageref{lastpage}}
\maketitle

\begin{abstract}
Small scale challenges suggest some missing pieces in our current understanding of dark matter. A cascade theory for dark matter is proposed to provide extra insights, similar to the cascade phenomenon in hydrodynamic turbulence. The kinetic energy is cascaded in dark matter from small to large scales involving a constant rate $\varepsilon_u$ ($\approx -4.6\times 10^{-7}m^2/s^3$). Confirmed by N-body simulations, the energy cascade leads to a two-thirds law for kinetic energy $v_r^2$ on scale $r$ such that $v_r^2 \propto (\varepsilon_u r)^{2/3}$. Equivalently, a four-thirds law can be established for halo density $\rho_s$ enclosed in the scale radius $r_s$ such that $\rho_s \propto \varepsilon_u^{2/3}G^{-1}r_s^{-4/3}$, which can be confirmed by galaxy rotation curves. Critical properties of dark matter might be obtained by identifying key constants on relevant scales. First, the largest halo scale $r_l$ can be determined by $-u_0^3/\varepsilon_u$, where $u_0$ is the velocity dispersion. Second, the smallest scale $r_{\eta}$ is dependent on the nature of dark matter. For collisionless dark matter, length scale $r_{\eta} \propto (-{G\hbar/\varepsilon_{u}}) ^{1/3}\approx 10^{-13}m$ was found along with the mass scale $m_{X}\propto (-{\varepsilon _{u} \hbar ^{5}}{G^{-4}})^{1/9} \approx 10^{12}GeV$, where $\hbar$ is the Planck constant. An uncertainty principle for momentum and acceleration fluctuations is also postulated. For self-interacting dark matter, $r_{\eta} \propto \varepsilon_{u}^2 G^{-3}(\sigma/m)^3$, where $\sigma/m$ is the cross-section of interaction. On halo scale, the energy cascade leads to an asymptotic density slope $\gamma=-4/3$ for fully virialized haloes with a vanishing radial flow, which might explain the nearly universal halo density. Based on the continuity equation, halo density is analytically shown to be closely dependent on the radial flow and mass accretion, such that simulated haloes can have different limiting slopes. A modified Einasto density profile is proposed accordingly.
\end{abstract}

\begin{keywords}
\vspace*{-10pt}
Dark matter flow; N-body simulations; Rotation curve; Collisionless; Self-interacting; Core-cusp; Energy cascade
\end{keywords}

\begingroup
\let\clearpage\relax
\tableofcontents
\endgroup
\vspace*{-20pt}

\section{Introduction}
\label{sec:1}
Standard CDM (cold dark matter) paradigm of cosmology has many successes in the formation and evolution of large scale structures and the contents and states of our universe \cite{Peebles:1984-Tests-of-cosmological-models,Spergel:2003-First-Year-Wilkinson-Microwave-Anisotropy,Komatsu:Seven-year-Wilkinson-Microwave-Anisotropy-Probe,Frenk:2012-Dark-matter-and-cosmic-structure}. Despite great successes, serious theoretical and observational difficulties still exist \cite{Perivolaropoulos:2022-Challenges-for,Bullock:2017-Small-Scale-Challenges-to-the}. Especially, CDM model predictions of structures on small scales (<1Mpc) are inconsistent with some observations. Examples are the core-cusp problem \cite{Flores:1994-Observational-and-Theoretical-Constraints,deBlok:2009-The-Core-Cusp-Problem}, the missing satellite problem \cite{Klypin:1999-Where-Are-the-Missing-Galactic, Moore:1999-Dark-Matter-Substructure}, the too-big-to-fail problem \cite{Boylan_Kolchin:2011-Too-big-to-fail,Boylan-Kolchin:2012-The-Milky-Ways-bright-satellites}. In addition, the origin of Baryonic Tully-Fishery relation (BTFR) and MOND (modified Newtonian dynamics) \cite{Milgrom:1983-A-Modification-of-the-Newtonia,McGaugh:2000-The-baryonic-Tully-Fisher-rela, Famaey:2013-Challenges-fo-CDM-and-MOND} is still not clear.

These small scale challenges might be related to each other \cite{Garrison-Kimmel:2014-Too-big-to-fail-in-the-Local-Group, Bullock:2017-Small-Scale-Challenges-to-the, Popolo:2014-A-unified-solution-to-the-small-scale} and suggest missing pieces in our current understandings. First, the cusp-core problem describes the discrepancy between the cuspy halo density predicted by cosmological CDM only N-body simulations and the cored density inferred from observational data for dwarf galaxies. The predicted halo density exhibits a cuspy profile with inner density $\rho(r) \propto r^{\gamma}$, where slope $\gamma$ persistently exceeds different observations \cite{Blok:2002GALAXIES:-STRUCTURE-GALAXIES,Blok:2003-Simulating-observations-of-dark-matter,Swaters:2002-The-Central-mass-distribution-in-dwarf,Naray:2011-Recovering-cores-and-cusps-in-dark-matter}. Even for the cuspy profile predicted by cosmological simulations, there seems no consensus on the exact value of asymptotic slope $\gamma$, but with a wide range between -1.0 to -1.5. Since the first prediction of $\gamma = -1.0$ in NFW profile \cite{Navarro:1997-A-universal-density-profile-fr}, the inner density slope of simulated haloes seems to have different values from $\gamma>-1.0$ \cite{Navarro:2010-The-diversity-and-similarity-of-simulated} to $\gamma=-1.2$  \cite{Diemand:2011-The-Structure-and-Evolution-of-Cold-Dark}, and $\gamma=-1.3$ \cite{Governato:2010-Bulgeless-dwarf-galaxies-and-dark-matter-cores,McKeown:2022-Amplified-J-factors-in-the-Galactic-Centre}. To summarize, some key questions are: is there an asymptotic slope for dark matter haloes? why there exists a nearly universal density profile? and why different inner slopes $\gamma$ exist in simulations?

The halo density inferred from observational data exacerbates the problem. Even the smallest predicted inner density slope from simulations is still greater than that from observations. Many solutions have been suggested to solve the cusp-core problem \cite{DelPopolo:2017-Small-scale-problems-of-the}. Within the CDM framework of collisionless dark matter, the baryonic solutions focus on different mechanisms for energy exchange between baryons and dark matter to enable a flatter inner density \cite{Navarro:1996-The-cores-of-dwarf-galaxy-haloes, Oh:2011-The-Central-Slope-of-Dark-Matter-Cores,Benitez:2019-Baryon-induced-dark-matter-cores}. 
Beyond the CDM framework, the self-interacting dark matter is proposed as a potential solution \cite{Spergel:2000-Observational-Evidence-for-Self-Interacting-Cold-Dark-Matter, Rocha:2013-Cosmological-simulations-with-self-interacting-dark-matter,Peter:2013-Cosmological-simulations-with-self-interacting-dark-matter}. The elastic scattering with a given cross-section facilitates the exchange of momentum and energy between dark matter particles and the formation of a flat core. Although the existence of dark matter is supported by numerous astronomical observations \cite{Rubin:1970-Rotation-of-Andromeda-Nebula-f,Rubin:1980-Rotational-Properties-of-21-Sc}, the nature and fundamental properties of dark matter are still a big mystery. No matter collisionless or self-interacting, some key questions remain open: what are the limiting length or density scales for dark matter if exist? what is the effect of self-interaction on these scales? what are the fundamental properties (particle mass, cross-section etc.) of dark matter? Answers to these questions would be critical for identifying and detecting dark matter.  

\begin{figure}
\includegraphics*[width=\columnwidth]{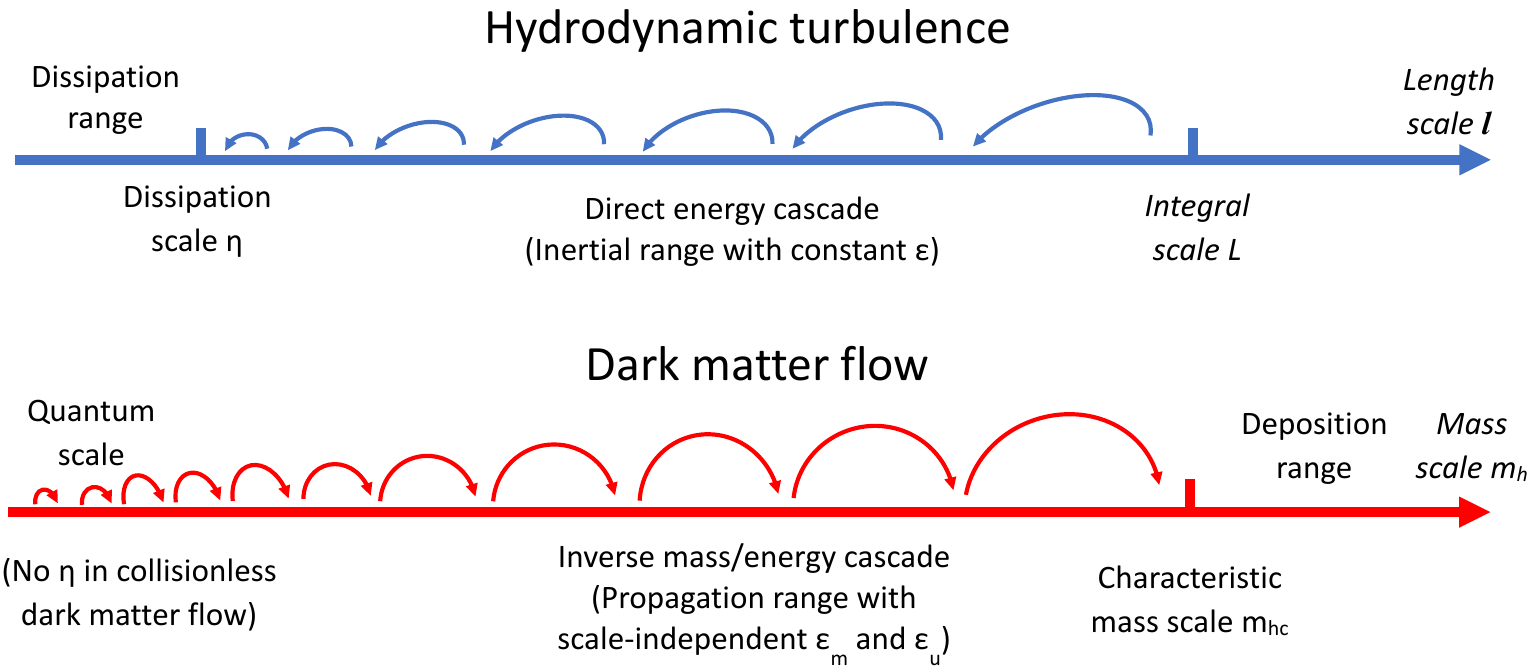}
\caption{Schematic plot of the direct energy cascade in turbulence and the inverse mass and energy cascade in dark matter flow. Haloes merge with single mergers to facilitate a continuous mass and energy cascade to large scales. Scale-independent mass flux $\varepsilon_{m}$ and energy flux $\varepsilon_{u}$ are expected for haloes smaller than a characteristic mass scale (propagation range similar to the inertial range in turbulence). Mass cascaded from small scales is consumed to grow haloes at scales above the characteristic mass (the deposition range similar to the dissipation range in turbulence), where mass and energy flux become scale-dependent \cite{Xu:2022-Dark_matter-flow-and-hydrodynamic-turbulence-presentation,Xu:2021-Inverse-mass-cascade-mass-function,Xu:2021-Inverse-and-direct-cascade-of-}.} 
\label{fig:S1}
\end{figure}

In this paper, a cascade theory for dark matter flow is proposed to provide some useful insights, similar to the cascade in hydrodynamic turbulence. Both dark matter flow and turbulence are typical non-equilibrium systems involving energy cascade as a key mechanism to continuously release energy and maximize system entropy. To grasp the key idea, we first present the cascade in turbulence that has been well-studied for many decades \cite{Taylor:1935-Statistical-theory-of-turbulan,de_Karman:1938-On-the-statistical-theory-of-i,Batchelor:1953-The-Theory-of-Homogeneous-Turb}. As shown in Fig. \ref{fig:S1}, turbulence consists of a collection of eddies (building blocks) on different length scale $l$ that are interacting with each other. The classical picture of turbulence is an eddy-mediated energy cascade process, where kinetic energy of large eddies feeds smaller eddies, which feeds even smaller eddies, and so on to the smallest scale $\eta$ where viscous dissipation is dominant. The direct energy cascade in turbulence can be best described by a poem \cite{Richardson:1922-Weather-Prediction-by-Numerica}: 
\smallbreak
\centerline{"Big whirls have little whirls, That feed on their velocity;}
\centerline{And little whirls have lesser whirls, And so on to viscosity."} 
\smallbreak

\noindent Despite the similarities, dark matter flow exhibits many different behaviors due to its collisionless and long-range interaction nature. First, unlike the turbulence that is incompressible on all scales, dark matter flow exhibits scale-dependent flow behaviors, i.e. a constant divergence flow on small scales and irrotational flow on large scales \cite{Xu:2023-On-the-statistical-theory-of-self-gravitating}. Second, the long-range gravity requires a broad spectrum of haloes to be formed to maximize the system entropy \citep{Xu:2023-Maximum-entropy-distributions-of-dark-matter}. In principle, haloes of different mass can be grouped into groups of haloes with the same mass $m_h$. Mass accretion facilitates a continuous mass and energy exchange between haloes groups on different mass scale $m_h$, i.e. an inverse mass and energy cascade (Fig. \ref{fig:S1}).   

The highly localized and over-dense haloes are a major manifestation of nonlinear gravitational collapse \cite{Neyman:1952-A-Theory-of-the-Spatial-Distri,Cooray:2002-Halo-models-of-large-scale-str} and the building blocks of dark matter flow, a counterpart to "eddies" in turbulence. The halo-mediated inverse mass cascade is not present in turbulence, but exists as a local, two-way, and asymmetric process in dark matter flow \cite{Xu:2021-Inverse-mass-cascade-mass-function}. The net mass transfer proceeds in a "bottom-up" fashion from small to large mass scales (inverse cascade) to allow for hierarchical structure formation. Haloes pass their mass onto larger and larger haloes, until halo mass growth becomes dominant over the mass propagation. From this description, mass cascade can be described by a similar poem with "eddies" (or "whirls") simply replaced by "haloes": 
\smallbreak
\centerline{"Little haloes have big haloes, That feed on their mass;} 
\centerline{And big haloes have greater haloes, And so on to growth."} 
\smallbreak

\noindent Energy cascade across halo groups is facilitated by the mass cascade and also a fundamental feature. Even on the halo scale, since haloes are non-equilibrium objects, energy cascade should also play a role in the abundance and internal structure of haloes \citep{Xu:2023-Dark-matter-halo-mass-functions-and}. In this paper, we focus on the energy cascade, its evidence from galaxy rotation curves, and its critical roles for halo internal structure and dark matter properties. 

\section{The constant rate of energy cascade}
\label{sec:2}
\begin{figure}
\includegraphics*[width=\columnwidth]{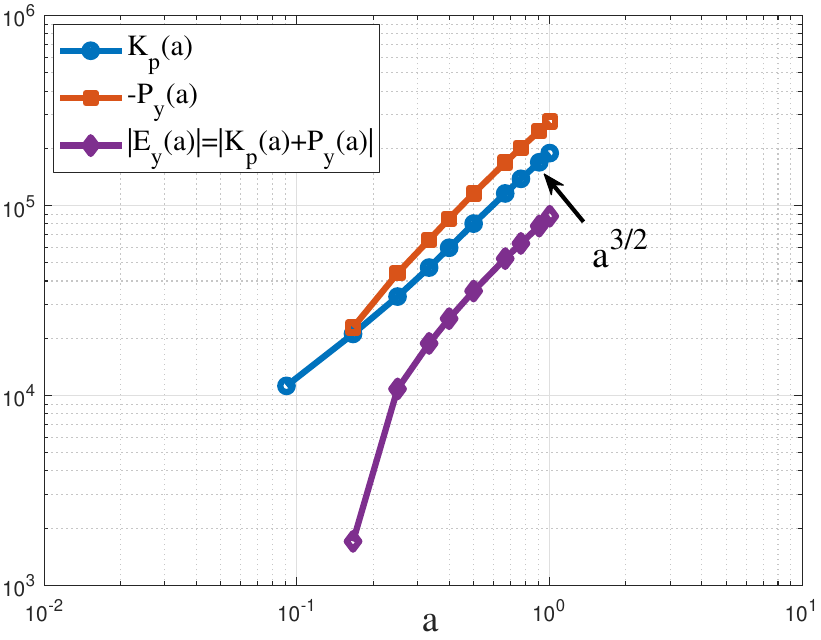}
\caption{The time variation of kinetic and potential energies from \textit{N}-body simulation. Both exhibit a power-law scaling with scale factor \textit{a}, i.e. $K_p$ and $P_y$ $\propto a^{3/2} \propto \varepsilon_u t$. The proportional constant $\varepsilon _{\boldsymbol{\mathrm{u}}}$ is estimated in Eq. \eqref{eq:2}.}
\label{fig:1}
\end{figure}

Particle-based N-body simulations are widely used to study the nonlinear gravitational collapse of dark matter \cite{Peebles:1980-The-Large-Scale-Structure-of-t}. The simulation data for this work was generated from N-body simulations by Virgo consortium \cite{Frenk:2000-Public-Release-of-N-body-simul,Jenkins:1998-Evolution-of-structure-in-cold}. One way to determine the constant rate of energy cascade $\varepsilon_u$ is from a cosmic energy equation for energy evolution of dark matter flow in expanding background \cite{Irvine:1961-Local-Irregularities-in-a-Univ,Layzer:1963-A-Preface-to-Cosmogony--I--The,Xu:2022-The-evolution-of-energy--momen},
\begin{equation} 
\label{eq:1} 
\frac{\partial E_{y} }{\partial t} +H\left(2K_{p} +P_{y} \right)=0,         
\end{equation} 
which is a manifestation of energy conservation. Here $K_{p}$ is the specific (peculiar) kinetic energy, $P_{y}$ is the specific potential energy in physical coordinate, $E_{y} =K_{p}+P_{y}$ is the total energy, $H={\dot{a}/a}$ is the Hubble parameter ($Ht=2/3$ for matter dominant universe), and \textit{a} is the scale factor. In statistically steady state, Eq. \eqref{eq:1} admits a linear solution of $K_{p} =-\varepsilon_u t$ and $P_{y} =1.4\varepsilon_u t$ (see Fig. \ref{fig:1}) such that $\varepsilon _{\boldsymbol{\mathrm{u}}}$ can be found as,  
\begin{equation} 
\label{eq:2} 
\varepsilon _{u} =-\frac{K_{p} }{t} =-\frac{3}{2} \frac{u^{2} }{t} =-\frac{3}{2} \frac{u_{0}^{2} }{t_{0} } =-\frac{9}{4} H_{0} u_{0}^{2} \approx -4.6\times 10^{-7} \frac{m^{2} }{s^{3} } ,     
\end{equation} 
where $u_{0} \equiv u\left(t=t_{0} \right)\approx 354.6{km/s} $ is the one-dimensional velocity dispersion of all dark matter particles and $t_{0}$ is the present age of universe (13.8 billion years). The constant $\varepsilon _{\boldsymbol{\mathrm{u}}}$ represents the rate of energy cascade across different scales. The negative value $\varepsilon _{\boldsymbol{\mathrm{u}}}<0$ reflects the direction (inverse) from small to large mass scales. 


\section{The 2/3 and -4/3 laws for energy and density}
\label{sec:3}
\begin{figure}
\includegraphics*[width=\columnwidth]{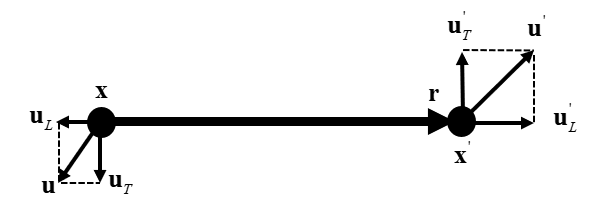}
\caption{Sketch of longitudinal and transverse velocities, where $\boldsymbol{\mathrm{u}}_{T} $ and $\boldsymbol{\mathrm{u}}_{T}^{'} $ are transverse velocities at two locations $\boldsymbol{\mathrm{x}}$ and $\boldsymbol{\mathrm{x}}^{'} $. $u_{L} $ and $u_{L}^{'} $ are two longitudinal velocities.} 
\label{fig:S2}
\end{figure}

To develop statistical theory of dark matter flow on all scales, different statistical measures can be introduced including the correlation, structure, dispersion functions, and power spectrum for density, velocity and potential fields \citep{Xu:2023-On-the-statistical-theory-of-self-gravitating}. Among different measures, structure functions are of particular interest that describes how energy is distributed and transferred across different length scales. For a pair of particles at two different locations $\boldsymbol{\mathrm{x}}$ and $\boldsymbol{\mathrm{x}}^{'}$ with velocity $\boldsymbol{\mathrm{u}}$ and $\boldsymbol{\mathrm{u}}^{'}$, the second order longitudinal structure function $S_{2}^{lp}$ (pairwise velocity dispersion in cosmology terms) is defined as
\begin{equation} 
\label{eq:3} 
S_{2}^{lp} \left(r,t\right)=\left\langle \left(\Delta u_{L} \right)^{2} \right\rangle =\left\langle \left(u_{L}^{'} -u_{L} \right)^{2} \right\rangle ,         
\end{equation} 
where $u_{L} =\boldsymbol{\mathrm{u}}\cdot \hat{\boldsymbol{\mathrm{r}}}$ and $u_{L}^{'} =\boldsymbol{\mathrm{u}}^{'} \cdot \hat{\boldsymbol{\mathrm{r}}}$ are two longitudinal velocities. The distance $r\equiv |\boldsymbol{\mathrm{r}}|=|\boldsymbol{\mathrm{x}}^{'} -\boldsymbol{\mathrm{x}}|$ and the unit vector $\hat{\boldsymbol{\mathrm{r}}}={\boldsymbol{\mathrm{r}}/r}$ (see Fig. \ref{fig:S2}). For a given scale \textit{r}, all particle pairs with the same separation \textit{r} can be identified in N-body simulation. The particle position and velocity data were recorded to compute the structure function in Eq. \eqref{eq:3} by averaging over all pairs with the same \textit{r} (i.e. a pairwise average).

In incompressible flow, the structure function has a small scale limit ${\mathop{\lim }\limits_{r\to 0}} S_{2}^{lp}=0$ because of $u_{L} \approx u_{L}^{'}$ due to the viscous force. However, in dark matter flow, ${\mathop{\lim }\limits_{r\to 0}} S_{2}^{lp} =2u^{2} \ne 0$ due to the collisionless nature \cite{Xu:2022-Postulating-dark-matter-partic,Xu:2022-Two-thirds-law-for-pairwise-ve}, where $u^2$ is velocity dispersion in Eq. \eqref{eq:2}. The pair of particles with a sufficiently small $r$ is more likely from the same halo, while different pairs can be from different haloes. Kinetic energy of particle pairs on scale $r$ includes contributions from both the relative motion of two particles and the motion of haloes that particle pair resides in. The kinetic energy from the motion of haloes ($2u^2$) is relatively the same for different pairs. Kinetic energy involved in the energy cascade should be the part due to the relative motion. Since the original structure function (pairwise dispersion) $S_{2}^{lp}(r)$ includes the total kinetic energy on scale $r$, a reduced structure function $S_{2r}^{lp}(r)=S_{2}^{lp}-2u^2$ can be introduced to take the common part out and include only the part from relative motion with the right limit ${\mathop{\lim }\limits_{r\to 0}} S_{2r}^{lp}=0$. This description indicates that $S_{2r}^{lp}(r)$ should be determined by and only by $\varepsilon _{u}$ ( ${m^{2} /s^{3}}$), gravitational constant $G$ ($m^3/kg\cdot s^2)$, and scale \textit{r}. By a simple dimensional analysis, this reduced structure function must follow a two-thirds law, i.e. $S_{2r}^{lp}(r) \propto \left(-\varepsilon _{u} \right)^{{2/3} } r^{{2/3}}$. 

\begin{figure}
\includegraphics*[width=\columnwidth]{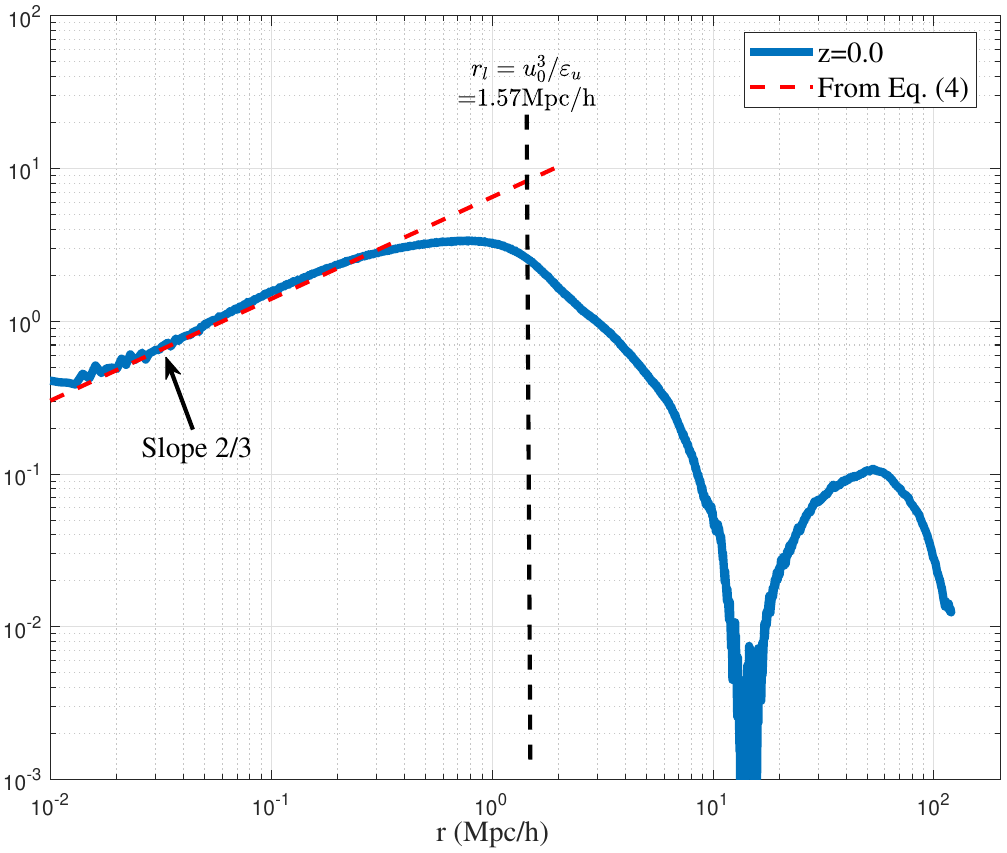}
\caption{The variation of reduced second order structure function $S_{2}^{lp}(r)$ with comoving scale $r$ at z=0. Structure function is normalized by the velocity dispersion $u^{2}$. A two-thirds law, i.e. $\propto \left(-\varepsilon _{u} \right)^{{2/3} } r^{{2/3}}$ can be clearly identified on small scale below a length scale $r_l=-u_0^3/\varepsilon_u$, where inverse energy cascade is established with a constant energy flux $\varepsilon_{u}<0$. The model from Eq. \eqref{eq:4} is also presented for comparison.}
\label{fig:4}
\end{figure}

\begin{figure}
\includegraphics*[width=\columnwidth]{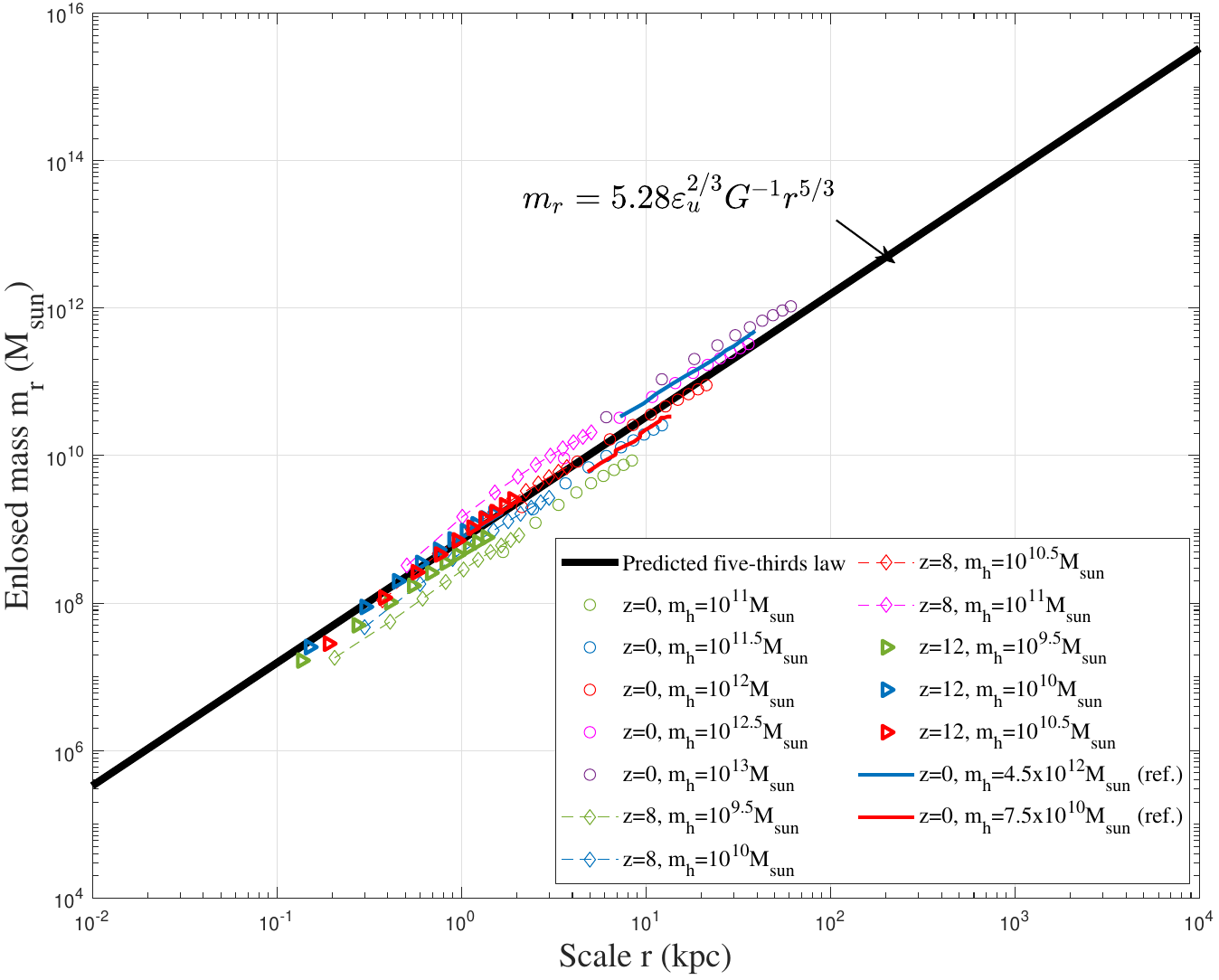}
\caption{The variation of enclosed mass $m_r$ with scale $r$ at different redshift from Illustris simulation. For all haloes with a given mass $m_h$, the average scale radius $r_s$ is calculated. The average enclosed mass $m_r$ is computed for different scale $r$ from $r=0.1r_s$ to $r=r_s$ with an increment of $0.1r_s$. Solid blue and red lines are results from ref. \cite{Zhao:2009-Accurate-Universal-Models-for-}. Both results confirm the predicted five-thirds law in Eq. \eqref{eq:7}.} 
\label{fig:5-2}
\end{figure}

Figure \ref{fig:4} plots the variation of $S_{2r}^{lp}$ with scale \textit{r} at z=0 from N-body simulations. The range $S_{2r}^{lp} \propto (-\varepsilon_ur)^{{2/3}}$ can be clearly identified below a critical length scale $r_l=-u_0^3/\varepsilon_u$. This range is formed due to the formation of haloes and inverse energy cascade. On small scale, $S_{2r}^{lp}$ (kinetic energy) should finally read
\begin{equation} 
\label{eq:4} 
\begin{split}
S_{2r}^{lp}\left(r\right)=a^{3/2}\beta_{2}^{*}{{\left( -{{\varepsilon }_{u}} \right)}^{2/3}}{r^{2/3}},     
\end{split}
\end{equation} 
where the proportional constant $\beta _{2}^{*} \approx 9.5$ can be found from Fig. \ref{fig:4}. Since $S_{2r}^{lp}$ represents the kinetic energy of relative motion on scale $r$, a different form of the two-thirds law (Eq. \eqref{eq:4}) can be obtained. By introducing a typical velocity $v_r$ on a given scale $r$,
\begin{equation} 
\label{eq:5} 
v_r^2 =S_{2r}^{lp}(r)/\left(2^{{2/3} }\beta_{2}^{*} a^{{3/2} } \right),     
\end{equation} 
the two-thirds law in Eq. \eqref{eq:4} can be equivalently written as,
\begin{equation} 
\label{eq:6} 
-\varepsilon_{u}=\frac{2v_r^{2}}{r}v_r =a_r v_r =\frac{2v_r^2}{r/v_{r}}= \frac{2v_r^3}{r},
\end{equation} 
where $a_r$ is the scale of acceleration. Equation \eqref{eq:6} also describes the cascade of kinetic energy in the inner halo region ($r<r_s$, where $r_s$ is the scale radius). The kinetic energy $v_{r}^2$ on scale $r$ is cascaded to large scale during a turnaround time of $t_r=r/v_r$. Combining Eq. \eqref{eq:6} with the virial theorem $Gm_r/r \propto v_r^2$ on scale $r$, we can easily obtain the typical mass $m_r$ (enclosed within $r$), density $\rho_r$, velocity $v_r$, and time $t_r$ on scale $r$, all determined by $\varepsilon_u$, $G$, and $r$:
\begin{equation} 
\label{eq:7} 
\begin{split}
&m_r = \alpha_r \varepsilon_u^{2/3}G^{-1}r^{5/3} \quad \textrm{and} \quad \rho_r = \beta_r \varepsilon_u^{2/3}G^{-1}r^{-4/3}, \\
&v_r \propto (-\varepsilon_u r)^{1/3} \quad \textrm{and} \quad t_r \propto (-\varepsilon_u)^{-1/3}r^{2/3},
\end{split}
\end{equation} 
where $\alpha_r$ and $\beta_r$ are two numerical constants. The predicted five-thirds law for mass $m_r$ enclosed in scale $r$ can be directly tested by N-body simulations. In this work, the large scale cosmological simulation Illustris (Illustris-1-Dark) was selected for comparison \cite{NELSON:2015-The-illustris-simulation}. Figure \ref{fig:5-2} presents the variation of enclosed mass $m_r$ with scale $r$ at different redshift $z$ for all haloes with a given mass $m_h$. Results from ref. \cite{Zhao:2009-Accurate-Universal-Models-for-} are also presented for comparison. Both results confirm the predicted five-thirds law in Eq. \eqref{eq:7}.
Next, the predicted four-thirds law $\rho_r(r) \propto r^{-4/3}$ for mean density enclosed in scale $r$ can be directly tested by data from galaxy rotation curves (see Fig. \ref{fig:5}).

\section{Halo density slope and mean radial flow}
\label{sec:4}
On small scale, inner halo region is assumed to be fully virialized for Eq. \eqref{eq:7} to be valid. A vanishing radial flow $u_{r}$ is expected from the stable clustering hypothesis, i.e. no net stream motion in physical coordinate along radial direction \cite{Mo:2010-Galaxy-formation-and-evolution, Xu:2023-On-the-statistical-theory-of-self-gravitating}. With $u_{r}\equiv 0$, there is no mass, momentum, and energy exchanges between different spherical shells. However, simulated haloes are non-equilibrium dynamic objects that might not be fully virialized, whose internal structure should be dependent on the radial flow. The relation we'll develop in this section might be useful for core/cusp controversy.

Since halo density models often involve a scale radius $r_s$ (the length scale of a halo), we may introduce a reduced spatial-temporal variable $x={r}/{r_s(t)}={c(t)r}/{r_h(t)}$, where $c\equiv c(t)$ is halo concentration and $r_h(t)$ is the virial size for halo of mass $m_h(t)$. Derivatives with respect to \textit{t} and \textit{r} can be derived using the chain rule,
\begin{equation}
\frac{\partial }{\partial t} =\frac{\partial }{\partial x} \frac{\partial x}{\partial t} =-\frac{x}{t} \frac{\partial \ln r_{s} }{\partial \ln t} \frac{\partial }{\partial x} \quad \textrm{and} \quad \frac{\partial }{\partial r} =\frac{\partial }{\partial x} \frac{\partial x}{\partial r} =\frac{1}{r_{s} } \frac{\partial }{\partial x}.    
\label{eq:8}
\end{equation}
A general function $F\left(x\right)$ can be introduced such that the mass $m_{r}(r,t)$ enclosed in radius \textit{r} and halo density $\rho _{h}(r,t)$ can all be expressed in terms of function $F\left(x\right)$, 
\begin{equation}
\begin{split}
&m_{r} \left(r,t\right)=m_{h} \left(t\right)\frac{F\left(x\right)}{F\left(c\right)},\\
&\rho _{h} \left(r,t\right)=\frac{1}{4\pi r^{2} } \frac{\partial m_{r} \left(r,t\right)}{\partial r} =\frac{m_{h} \left(t\right)}{4\pi r_{s}^{3} } \frac{F^{'} \left(x\right)}{x^{2} F\left(c\right)}.   
\end{split}
\label{eq:9}
\end{equation}
\noindent The time derivative of $\rho _{h} \left(r,t\right)$ can be obtained from Eq. \eqref{eq:9},
\begin{equation} 
\label{eq:10} 
\frac{\partial \rho _{h} \left(r,t\right)}{\partial t} =\frac{1}{4\pi r^{2} } \frac{\partial ^{2} m_{r} \left(r,t\right)}{\partial r\partial t} .        
\end{equation} 
The mass continuity equation for a spherical halo simply reads,
\begin{equation} 
\label{eq:11} 
\frac{\partial \rho _{h} \left(r,t\right)}{\partial t} +\frac{1}{r^{2} } \frac{\partial \left[r^{2} \rho _{h} \left(r,t\right)u_{r} \left(r,t\right)\right]}{\partial r} =0,       
\end{equation} 
where $u_{r} \left(r,t\right)$ is the mean radial flow velocity. From Eqs. \eqref{eq:10} and \eqref{eq:11}, the enclosed mass $m_{r}$ is related to the radial flow as
\begin{equation} 
\label{eq:12} 
\frac{\partial m_{r} \left(r,t\right)}{\partial t} =-4\pi r^{2} u_{r} \left(r,t\right)\rho _{h} \left(r,t\right).   \end{equation} 
With $m_{r}$ and $\rho _{h}$ from Eq. \eqref{eq:9}, the radial flow $u_r$ simply reads
\begin{equation} 
\label{eq:13} 
u_{r} =-\frac{1}{4\pi r^{2} } \frac{\partial \ln m_{r} }{\partial \ln t} \frac{m_{r} \left(r,t\right)}{\rho _{h} \left(r,t\right)t} =-\frac{r_{s} \left(t\right)}{t} \frac{\partial \ln m_{r} }{\partial \ln t} \frac{F\left(x\right)}{F^{'} \left(x\right)} .     
\end{equation} 
While from Eq. \eqref{eq:9} for $m_{r}$, we have
\begin{equation} 
\label{eq:14} 
\frac{\partial \ln m_{r} }{\partial \ln t} =\frac{\partial \ln m_{h} }{\partial \ln t} -\frac{xF^{'} \left(x\right)}{F\left(x\right)} \frac{\partial \ln r_{s} }{\partial \ln t} -\frac{cF^{'} \left(c\right)}{F\left(c\right)} \frac{\partial \ln c}{\partial \ln t} .       
\end{equation} 
Substituting Eq. \eqref{eq:14} into Eq. \eqref{eq:13}, the normalized radial flow $u_h$ reads
\begin{equation} 
\label{eq:15} 
u_h(x) = u_{r}\frac{t}{r_s}=\left[x\frac{\partial \ln r_{s} }{\partial \ln t} +\left(\frac{\partial \ln F\left(c\right)}{\partial \ln t} -\frac{\partial \ln m_{h} }{\partial \ln t} \right)\frac{F\left(x\right)}{F^{'} \left(x\right)} \right].     
\end{equation} 
For fast growing haloes in their early stage with a constant concentration $c=3.5$ and scale radius $r_s(t) \propto m_h(t) \propto t$, using Eq. \eqref{eq:15}, the cored density profiles (pISO, Einasto, etc. with $F(x)\propto x^3$ in Eqs. \eqref{eq:18} and \eqref{eq:19}) lead to $u_h = 2x/3$ for small $x$, while cuspy density profile (NFW with $F(x)\propto x^2$ in Eq. \eqref{eq:19}) leads to $u_h = x/2$ for small $x$. Taking the derivative in Eq. \eqref{eq:15} and combining with Eq. \eqref{eq:9}, the density slope $\gamma$ can be obtained exactly as,
\begin{equation} 
\label{eq:16} 
\gamma = \frac{\partial \ln \rho_h}{\partial \ln x}=\frac{\frac{\partial u_h}{\partial x}+\frac{\partial \ln  m_r(r_s,t)}{\partial \ln t}-\frac{\partial \ln r_s}{\partial \ln t}}{\frac{\partial \ln r_s}{\partial \ln t}-\frac{u_h}{x}}-2.
\end{equation} 
Clearly, the spatial variation of slope $\gamma$ comes from the radial flow $u_h$, while the time variation of $\gamma$ comes from $r_s(t)$ and $m_r(r_s,t)$ due to mass accretion. For fully virialized haloes or the virialized inner core, we should have $u_h \equiv 0$ such that the asymptotic slope $\gamma$ reads
\begin{equation} 
\label{eq:17} 
\gamma = \frac{\partial \ln \rho_h}{\partial \ln x}={\frac{\partial \ln m_r(r_s,t)}{\partial \ln r_s}}-3.
\end{equation} 

On halo scale, energy cascade with a constant rate $\varepsilon_u$ is valid for all scales $r\le r_s$. Taking the enclosed mass $m_r(r_s)$ as the mass scale in Eq. \eqref{eq:7}, we found $m_r(r_s) \propto r_s^{5/3}$. For fully virialized haloes with $u_h\equiv 0$, slope $\gamma=-4/3$ or a cuspy density $\rho_h(r) \propto r^{-4/3}$ can be obtained from Eq. \eqref{eq:16}. Therefore, fully virialized haloes should have universal cuspy density profiles due to energy cascade. In other words, simulated haloes might have different slope $\gamma$ due to nonzero radial flow and different mass accretion rate (Eq. \eqref{eq:16}). 

The baryonic feedback provides potential mechanisms to enhance the gradient of $u_h$ (deformation rate) in Eq. \eqref{eq:16}) and flatten the inner density. \rev{To better illustrate, we can approximate Eq. \eqref{eq:16} in the core region as:
\begin{equation} 
\label{eq:16-2} 
\begin{split}
&\gamma \approx \frac{\frac{\partial \ln m_r(r_s)}{\partial \ln r_s}}{1-\mu}-3, \quad  \mu = \frac{\partial u_h}{\partial x}\bigg /\frac{\partial ln r_s}{\partial ln t} \approx \frac{u_h}{x}\bigg /\frac{\partial ln r_s}{\partial ln t}.
\end{split}
\end{equation} 
Here dimensionless number $\mu$ represents the competition between radial flow and halo mass accretion, i.e. the deformation rate normalized by the rate of change in halo core size. The exponent $m={\partial ln r_s}/{\partial ln t}$ decreases with time and increases with halo size to $m \approx 1$ for large haloes. From scaling laws for $m_r$ in Eq.\eqref{eq:7}, $m_r(r_s) \propto r_s^{5/3}$. From Eq. \eqref{eq:16-2}, strong supernova-driven outflows in interstellar medium may lead to a non-zero radial outflow $u_h$ of dark matter or a greater $\mu$ such that core density becomes flatter from Eq. \eqref{eq:16-2}. This effect should be greater in smaller haloes due to smaller exponent $m$. Haloes should also be large enough for sufficient fraction of baryons converted into stars to allow feedback. Density cores are only developed in a certain range mass of haloes ($10^{10}$ to $10^{11} M_{\odot}$), as confirmed by cosmological simulations \cite{Chan:2015-The-impact-of-baryonic-physics-on-the-structure}. }

Finally, since $u_h(x=0)\equiv 0$, there exists an asymptotic slope $\gamma$ for simulated haloes for small $x\rightarrow 0$ that is dependent on the local gradient of $u_h$ around $x=0$ (see Eq. \eqref{eq:16}). Therefore, a better density profile can be proposed for simulated haloes (see \cite{Xu:2023-Dark-matter-halo-mass-functions-and} for details)
\begin{equation} 
\label{eq:18} 
\begin{split}
&\rho_h(r) = \rho_0 x^\gamma \exp\left(-\frac{2}{\alpha}x^{\alpha}\right),\\
&F(x) = \Gamma(\frac{3+\gamma}{\alpha})-\Gamma(\frac{3+\gamma}{\alpha},\frac{2}{\alpha}x^{\alpha}).
\end{split}
\end{equation} 
Here $\rho_0$ and $\alpha$ are density and shape parameters and slope $\gamma<0$. Different simulated haloes might have different slope $\gamma$ due to different behavior of radial flow $u_h$ at small $x$ and halo mass accretion.   

\section{Testing -4/3 law against rotation curves}
\label{sec:5}
Next, the predicted four-thirds law in Eq. \eqref{eq:7} ($\rho_r(r_s) \propto r_s^{-4/3}$) is tested against galaxy rotation curves that contain important information for dark matter haloes. In practice, rotational curves can be first decomposed into contributions from different mass components. Model parameters for halo density (scale radius $r_s$ and density scale $\rho_0$) are obtained by fitting to the decomposed rotation curve. In this work, we use three sources of galaxy rotation curves,
\begin{enumerate}
\item \noindent SPARC (Spitzer Photometry \& Accurate Rotation Curves) including 175 late-type galaxies \cite{Lelli:2016-SPARC-Mass-Models-for-175-Disk-Galaxies,Li:2020-A-Comprehensive-Catalog-of-Dark-Matter-Halo-Models};
\item \noindent DMS (DiskMass Survey) including 30 spiral galaxies \cite{Martinsson:2014-The-DiskMass-Survey};
\item \noindent SOFUE (compiled by Sofue) with 43 galaxies \cite{Sofue:2016-Rotation-curve-decomposition-f}.
\end{enumerate}

\rev{The mass modeling of galaxies is always challenging because of the uncertainties in the stellar mass-to-light ratio \textgamma. Both the DiskMass Survey (DMS) and stellar population synthesis models (SPS) suggest an almost constant \textgamma in near-infrared band for galaxies of different masses and morphologies \cite{Lelli:2016-SPARC-Mass-Models-for-175-Disk-Galaxies}. Wang and Chen removed the SPARC galaxies that have a bulge component to improve the fitting quality because of different mass-to-light ratio between the bulge and the disk \cite{Wang:2021-Comparison-of-Modeling-SPARC-spiral-galaxies}. For SPARC sample used in this work, 32 galaxies with significant bulges adopt \textgamma$_{bul}$ = 1.4 \textgamma$_{disk}$ as suggested by SPS models.} 

For pseudo-isothermal (pISO) and NFW density, we have
\begin{equation} 
\label{eq:19} 
\begin{split}
&\rho_{pISO}=\frac{\rho_0}{1+x^2}, \quad F(x) = x-\textrm{arctan}(x) \approx \frac{x^3}{3}, \\
&\rho_{NFW}=\frac{\rho_0}{x(1+x)^2},\quad F(x) = \textrm{log}(1+x)-\frac{x}{(1+x)} \approx \frac{x^2}{2},
\end{split}
\end{equation} 
where $\rho_0$ is a density parameter. From Eq. \eqref{eq:9}, halo density at $r_s$ is 
\begin{equation} 
\label{eq:20} 
\rho _{h}(r_s)=\rho _{h}(x=1)=\bar\rho_h\frac{c^3F'(1)}{3F(c)}=\Delta_c\rho_{crit}\frac{c^3F'(1)}{3F(c)},      
\end{equation} 
where $\bar\rho_h$ is the mean halo density. In this work, $\Delta_c=200$ and critical density $\rho_{crit}=3H_0^2/8\pi G=10^{-26}kg/m^3$. Using Eqs. \eqref{eq:19} and \eqref{eq:20}, concentration $c$ can be obtained from fitted model parameter  $\rho_0$. The mean density within $r_s$ (density scale $\rho_s$) now reads 
\begin{equation} 
\label{eq:21} 
\rho _{s}(r_s) =\frac{m_r(r_s)}{4\pi r_s^3/3}=\frac{F(1)c^3}{F(c)}\bar\rho_h=\frac{F(1)c^3}{F(c)}\Delta_c\rho_{crit},      
\end{equation} 

\begin{figure}
\includegraphics*[width=\columnwidth]{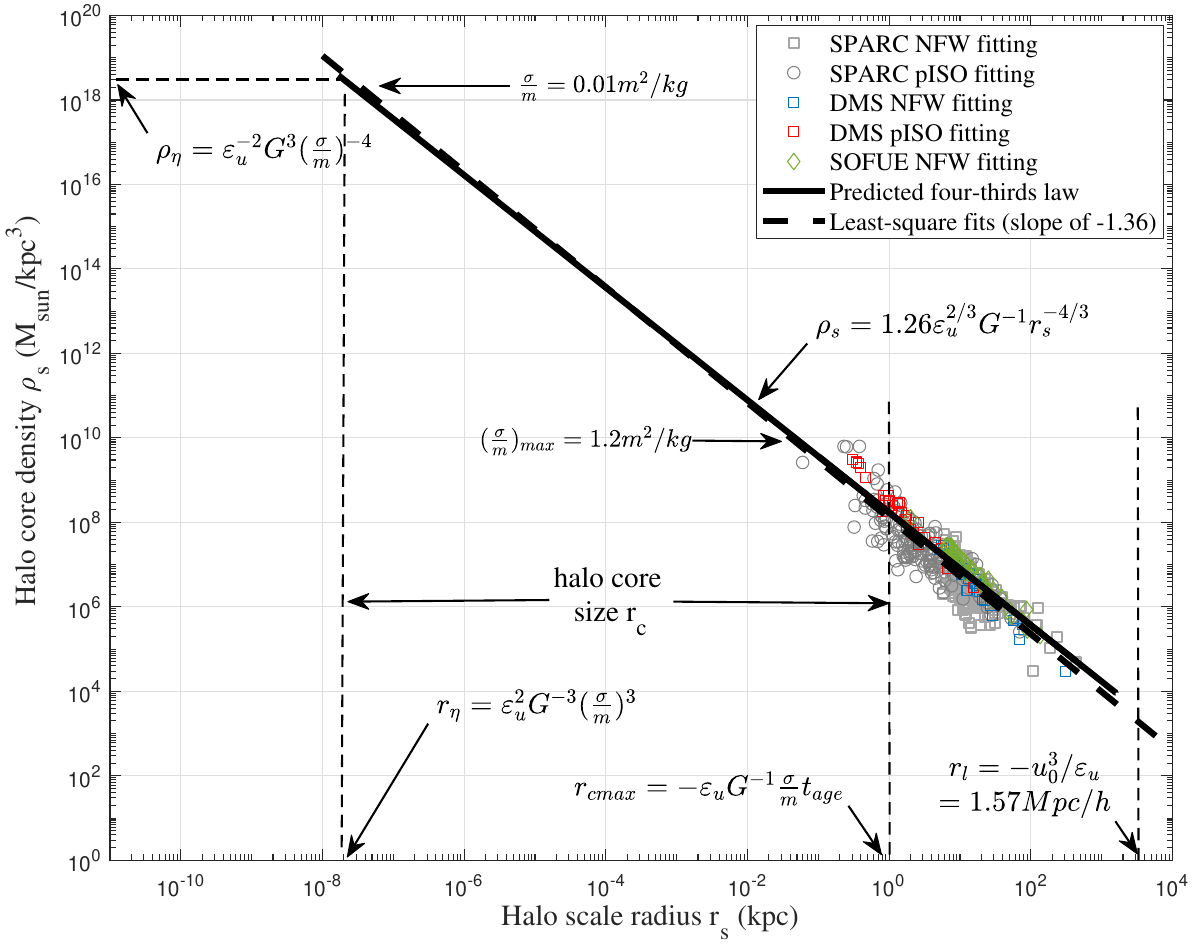}
\caption{The predicted -4/3 law tested against actual data from galaxy rotation curves. Good agreement confirms the existence of inverse energy cascade with a constant rate $\varepsilon_u$. The self-interacting dark matter model should modify the lowest size $r_{\eta}$ and maximum density $\rho_{\eta}$ determined by the cross-section $\sigma/m$, below which no coherent structure can exist. The largest possible core size $r_{cmax}$ due to self-interaction is determined by the age of haloes $t_{age}$. The largest scale $r_l$ is determined by the velocity dispersion $u_0$ and $\varepsilon_u$.}
\label{fig:5}
\end{figure}

\rev{Figure \ref{fig:5} presents the variation of typical density $\rho_s$ with scale $r_s$ obtained from three different sources of galaxy rotation curves. Strong correlation exists between core density and scale radius with a Pearson correlation coefficient of -0.91. The four-thirds law (Eq. \eqref{eq:7}) is also plotted for comparison with coefficients $\beta_r=1.26$ and $\alpha_r=5.28$ obtained from these data. The black dash line shows the least-square fit to data from all 248 galaxies with the best slope of -1.36$\pm0.05$ that is very close to -4/3. The R-square value (the percentage of data variation that can be explained by model) of the fit is 0.82 with root-mean-square scatter of all data around 0.3dex. Previous studies for the scaling between halo central density and core radius have different slopes for different model fits. A constant surface density ($\rho_s r_s$=const or $\rho_s \propto r_s^{-1.04}$) was first noticed from 55 rotation curves of spiral galaxies \cite{Kormendy:2004-Dark-Matter-in-Galaxies} for isothermal (ISO) fits. While slope becomes -1.20 for pseudo-isothermal (pISO) fits in the same study. Spano et al. also found a slope of -0.93 for ISO fits and -1.0 for NFW fits based on the data from 36 spiral galaxies with central density across three orders \cite{Spano:2008-GHASP-an-kinematic-survey-spiral}. Some studies also suggest a non-constant surface density ($\rho_s r_s\neq$const). Examples are the slope of -2 \cite{Barnes:2004-Mass-Models-for-Spiral-Galaxies} based on 40 spiral galaxies with core density across three orders. A slope of -1.46 was also found in galaxy cluster \cite{Chan:2014-A-tight-scaling-relation} that is close to this work.}

In this work, we proposed a theory for the existence of such scaling laws. Dark matter haloes obtained from rotation curves follow the predicted four-thirds law across 6 orders in both size and density. Equivalently, the inner cuspy density $\rho_h \propto r^{-4/3}$ for virialized haloes with a vanishing radial flow is also confirmed by this plot or from Eq. \eqref{eq:17}. This plot also confirms the existence of a constant rate of cascade $\varepsilon_u$ below the largest halo scale $r_l$. Other relevant quantities (density, pressure, energy etc.) on scale $r_l$ are similarly determined by constants $\varepsilon_u$, $G$, and $u_0$ \cite{Xu:2022-The-origin-of-MOND-acceleratio}. 

\begin{figure}
\includegraphics*[width=\columnwidth]{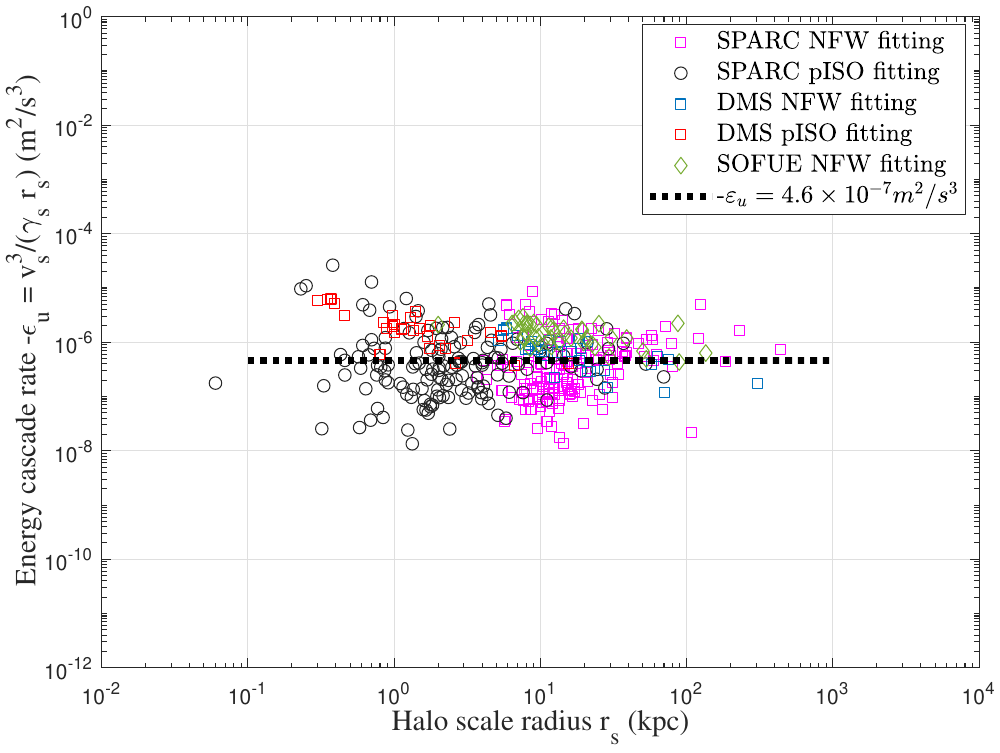}
\caption{The constant rate of inverse energy cascade $\varepsilon_u$ from galaxy rotation curve data. The dispersion in data might come from the spatial intermittence of energy cascade. Dwarf galaxy tends to have smaller $\varepsilon_u$, consistent with its density profile \citep{Xu:2023-Dark-matter-halo-mass-functions-and}.}
\label{fig:6}
\end{figure}
Finally, we can choose the circular velocity at $r_s$ as the typical velocity $v_s=\sqrt{Gm_r(r_s)/r_s}$ such that $-\varepsilon_u=v_s^3/(\gamma_s r_s)$ (from Eq. \eqref{eq:7}). Figure \ref{fig:6} presents the rate of energy cascade $\varepsilon_u$ obtained from three sources of galaxy rotation curves with $\gamma_s = 6.83$. The dispersion in data might come from the spatial intermittence of energy cascade such that haloes in different local environment may have slightly different $\varepsilon_u$.

\section{Scales for self-interacting dark matter}
\label{sec:6}
To solve the core-cusp problem, one option is the self-interacting dark matter (SIDM) model, where the cascade theory can be used to determine relevant scales. The cross-section $\sigma/m$ of self-interaction should introduce additional scales $r_{\eta}$, $\rho_{\eta}$, and $m_\eta$ (see Fig. \ref{fig:5}), beyond which no structure can be formed due to self-interaction. These scales can be obtained by requiring at least one scatter per particle during the typical time $t_r$, i.e. $\rho_r (\sigma/m) v_r t_r=1$ in Eq. \eqref{eq:7}. Combine this with the virial theorem and constant energy cascade in Eq. \eqref{eq:6}, the length scale $r_{\eta}$, density scale$\rho_{\eta}$, and mass scale $m_{\eta}$ are all determined by $\varepsilon_u$, $G$, and $\sigma/m$ (see Fig. \ref{fig:5} and Eq. \eqref{eq:22-2}),
\begin{equation} 
\begin{split}
&r_{\eta} =\varepsilon_{u}^{2}G^{-3}(\sigma/m)^{3}, \quad m_{\eta} =\varepsilon_{u}^4G^{-6}(\sigma/m)^{5},\\
&\rho_{\eta} =\varepsilon_{u}^{-2}G^{3}(\sigma/m)^{-4}
\end{split}
\label{eq:22-2} 
\end{equation}
From Eq. \eqref{eq:22-2}, the upper limit of cross section can be estimated as,
\begin{equation} 
\begin{split}
&\frac{\sigma}{m} \le \left(\frac{r_{\eta}}{1kpc}\right)^{\frac{1}{3}} \frac{G(1kpc)^{1/3}}{\varepsilon_{u}^{2/3}}=35.1\frac{cm^2}{g} \left(\frac{r_{\eta}}{1kpc}\right)^{\frac{1}{3}}.
\end{split}
\label{eq:22-3} 
\end{equation}
From galaxy rotation curves with a maximum core density around $\rho_{\eta} \approx 10^{10} M_{sun}/(kpc)^3$ and $r_{\eta} \approx 0.04$kpc (Fig. \ref{fig:5}), we can safely estimate the upper limit of $\sigma/m\le 12cm^2/g$ from Eq. \eqref{eq:22-3}. High resolution rotation curves for dwarf galaxies should provide more stringent constraints on the cross-section by determining the smallest length scale $r_{\eta}$. \rev{For comparison, the first constraint using colliding galaxy clusters found that $\sigma/m<5cm^2/g$ \cite{Markevitch:2004-Direct-Constraints-on-the-Dark-Matter}. Analysis performed on the Bullet Cluster leads to an upper limit of $\sigma/m<4cm^2/g$ from MACS J0025.4-1222 \cite{Brada:2008-Revealing-the-Properties-of-Dark-Matter} and $\sigma/m<7cm^2/g$ from DLSCL J0916.2+2951 \cite{Dawson:2012-Discovery-of-a-Dissociative-Galaxy}. More stringent constraint was also obtained from Bullet Clusters 1E 0657-56 with  $\sigma/m<2cm^2/g$ \cite{Robertson:2016-What-does-the-Bullet-Cluster-tell}.} 

In addition, due to self-interaction, dark matter halo should have an isothermal core with a maximum core size $r_{cmax}$ by requiring at least one scatter during the age of haloes ($t_{age}\approx 1/H_0$), i.e. $\rho_r (\sigma/m) v_r t_{age}=1$ such that (using Eq. \eqref{eq:7})
\begin{equation} 
\label{eq:22} 
\frac{r_{cmax}}{\sigma/m}=\varepsilon_u G^{-1}t_{age}=10kpc\frac{g}{cm^2}.
\end{equation} 
For cross-section $\sigma/m=0.01m^2/kg=0.1cm^2/g$ used in SIDM cosmological simulations \cite{Rocha:2013-Cosmological-simulations-with-self-interacting-dark-matter}, halo core size is between $r_{\eta}$ and the maximum core size $r_{cmax} \approx 1kpc$. In other words, maximum core size $r_{cmax}$ formed from self-interaction (not any other mechanisms) can be used to identify the cross-section of postulated self-interaction. 

In hydrodynamic turbulence, the smallest length scale $\eta=(\nu^3/\varepsilon)^{1/4}$ \cite{Kolmogoroff:1941-Dissipation-of-energy-in-the-l} is determined by fluid viscosity $\nu$ and rate of energy cascade $\varepsilon$. The kinetic energy is injected at large scale and cascaded down to small scales. Below scale $\eta$, structures (eddies) are destroyed by viscous force and kinetic energy is dissipated into heat to increase system entropy (Fig. \ref{fig:S1}). Here $\varepsilon$ can be a variable manually controlled or adjusted by the rate of energy injection on large scale. For faster mixing, thinking about stirring the coffee-milk fluid harder (the larger $\varepsilon$), the scale $\eta$ would be smaller and mixing would be faster due to greater velocity on that scale. However, for dark matter flow in our universe, inverse (NOT direct) energy cascade is required for hierarchical structure formation on large scales. The rate of energy cascade $\varepsilon_u$ is a constant on small scales $r<r_l$ that cannot be manually controlled, which should be a fundamental constant relevant to dark matter properties. Finally, for self-interacting dark matter, it should be interesting to identify the smallest and greatest halo core size ($r_{\eta}$ and $r_{cmax}$) to constrain the cross-section using Eqs. \eqref{eq:22-3} and \eqref{eq:22}. 

\section{Scales for collisionless dark matter}
\label{sec:7}
\begin{figure}
\includegraphics*[width=\columnwidth]{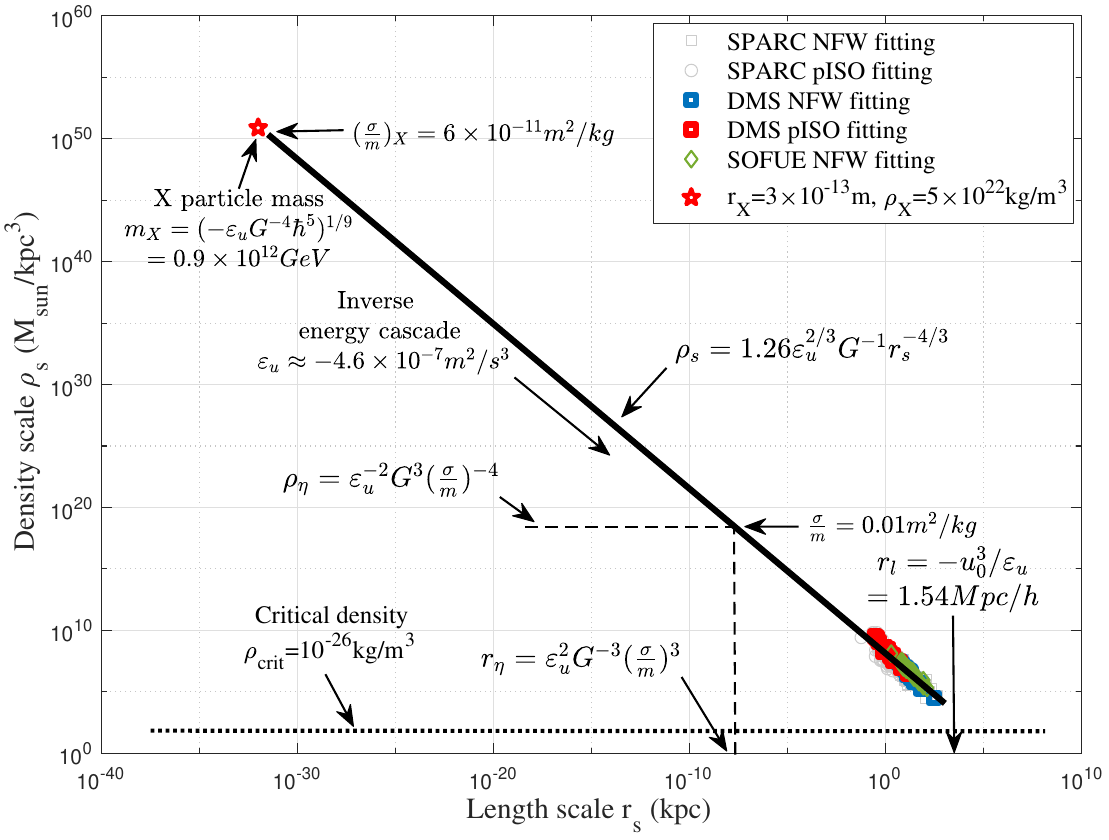}
\caption{For fully collisionless dark matter, we may extend the predicted -4/3 law in Fig. \ref{fig:5} to the smallest scale where quantum effect can be important (red star). On this scale, dark matter particles properties are determined by three constants: $\varepsilon_u$, $G$, and Planck constant $\hbar$ \cite{Xu:2022-Postulating-dark-matter-partic}.}
\label{fig:7}
\end{figure}

Now let's consider the second option: dark matter is fully collisionless with flat halo core formed by baryonic feedback mechanism. In this scenario, due to the collisionless nature and scale-independence of $\varepsilon_u$, the four-thirds law should extend from galaxy scale to the smallest scale where the quantum effect becomes important (Fig. \ref{fig:7}). This extension is more than 30 orders in size, which hopefully allows us to predict the mass, size and properties of dark matter particles (the $X$ particle) from three basic constants, i.e. $\varepsilon_u$, $G$, and Planck constant $\hbar$ \cite{Xu:2022-Postulating-dark-matter-partic}. Two examples are the critical mass and length scales,
\begin{equation}
\begin{split}
m_{X} \propto \left(-\frac{\varepsilon _{u} \hbar ^{5}}{G^{4}} \right)^{\frac{1}{9}} \approx 10^{12}GeV, \quad
r_{X} \propto \left(-\frac{G\hbar}{\varepsilon _{u}} \right)^{\frac{1}{3}} \approx 10^{-13}m. 
\end{split}
\label{eq:23}
\end{equation}
If this is true, the constant $\varepsilon_u$ might be an intrinsic property of dark matter with a similar origin as Planck constant $\hbar$ for two reasons:
\begin{enumerate}
\item \noindent For fully collisionless dark matter, there exists a unique "symmetry" between position and velocity in phase space. At any given location, collisionless particles can have multiple values of velocity (multi-stream regime). Similarly, particles with same velocity can be found at different locations. This "symmetry" in phase space is not possible for any non-relativistic baryonic matter. 
\item \noindent Due to the long-rang gravitational interaction, there exist fluctuations (uncertainty) not only in position ($\textbf{x}$) and velocity ($\textbf{v}=\dot{\textbf{x}}$), but also in acceleration ($\textbf{a}=\dot{\textbf{v}}$) \cite{Xu:2022-The-origin-of-MOND-acceleratio}. 
\end{enumerate}
With $\psi(x)$, $\varphi (p)$, and $\mu (a)$ as wave functions for position, momentum, and acceleration, we can write
\begin{equation}
\label{eq:24}
\begin{split}
&\psi(x)={\frac {1}{\sqrt {2\pi \hbar }}\int _{-\infty }^{\infty }\varphi (p)\cdot e^{ip\cdot x/\hbar }\,dp~,} \\
&\varphi(p)={\frac {1}{\sqrt {2\pi \mu_X }}\int _{-\infty }^{\infty }\mu (a)\cdot e^{ia\cdot p/\mu_X }\,da~,} 
\end{split}
\end{equation}
where constant $\mu_X=-m_X\varepsilon_u=7.44\times10^{-22}kg\cdot m^2/s^3$. An energy scale $\sqrt{\hbar\mu_X}\approx 10^{-9}eV$ can be obtained for the possible dark radiation due to dark matter annihilation or decay. Here we have two pairs of conjugate variables: i) position $x$ and momentum $p$, and ii) momentum $p$ and acceleration $a$. By following the standard wave mechanics (will not repeat here), two uncertainty principles can be established for fluctuations of position, momentum, and acceleration for collisionless dark matter, 
\begin{equation}
\label{eq:25}
\sigma_x\sigma_p\ge\hbar/2 \quad \textrm{and}  \quad \sigma_p\sigma_a\ge \mu_X/2.
\end{equation}
More experiment data might be required to test this postulation.

\section{Conclusions}
\label{sec:8}
Small scale challenges suggest some missing pieces in our current understanding of dark matter. A cascade theory for dark matter flow provides extra insights. The energy cascade with a constant rate $\varepsilon_u$ across different scales is a fundamental feature of dark matter flow. N-body simulation suggests a two-thirds law,  i.e. the kinetic energy $v_r^2 \propto (\varepsilon_u r)^{2/3}$ on scale $r$. This is equivalent to a four-thirds law for density on the same scale, i.e. $\rho_r \propto \varepsilon_u^{2/3}G^{-1}r^{-4/3}$, which can be directly confirmed by data from N-body simulations and galaxy rotation curves. By identifying key constants on relevant scales, limiting scales for collisionless (determined by $\varepsilon_u$, $G$, $\hbar$) or self-interacting dark matter (by $\varepsilon_u$, $G$, $\sigma/m$) might be obtained. On halo scale, based on the continuity equation, halo density is shown to be closely dependent on the radial flow and mass accretion. The asymptotic density slope $\gamma=-4/3$ can be obtained for fully virialized haloes with a vanishing radial flow. Simulated haloes can have different limiting slopes due to finite radial flow and different rates of mass accretion. The baryonic feedback might enhance the radial flow and flatten the core density. A modified Einasto density profile is proposed accordingly. 


\section*{Data Availability}
Two datasets for this article, i.e. a halo-based and correlation-based statistics of dark matter flow, are available on Zenodo \cite{Xu:2022-Dark_matter-flow-dataset-part1,Xu:2022-Dark_matter-flow-dataset-part2}, along with the accompanying presentation "A comparative study of dark matter flow \& hydrodynamic turbulence and its applications" \cite{Xu:2022-Dark_matter-flow-and-hydrodynamic-turbulence-presentation}.

\bibliographystyle{Papers}
\bibliography{Papers}

\begin{thebibliography}{76}%
\makeatletter
\providecommand \@ifxundefined [1]{%
 \@ifx{#1\undefined}
}%
\providecommand \@ifnum [1]{%
 \ifnum #1\expandafter \@firstoftwo
 \else \expandafter \@secondoftwo
 \fi
}%
\providecommand \@ifx [1]{%
 \ifx #1\expandafter \@firstoftwo
 \else \expandafter \@secondoftwo
 \fi
}%
\providecommand \natexlab [1]{#1}%
\providecommand \enquote  [1]{``#1''}%
\providecommand \bibnamefont  [1]{#1}%
\providecommand \bibfnamefont [1]{#1}%
\providecommand \citenamefont [1]{#1}%
\providecommand \href@noop [0]{\@secondoftwo}%
\providecommand \href [0]{\begingroup \@sanitize@url \@href}%
\providecommand \@href[1]{\@@startlink{#1}\@@href}%
\providecommand \@@href[1]{\endgroup#1\@@endlink}%
\providecommand \@sanitize@url [0]{\catcode `\\12\catcode `\$12\catcode
  `\&12\catcode `\#12\catcode `\^12\catcode `\_12\catcode `\%12\relax}%
\providecommand \@@startlink[1]{}%
\providecommand \@@endlink[0]{}%
\providecommand \url  [0]{\begingroup\@sanitize@url \@url }%
\providecommand \@url [1]{\endgroup\@href {#1}{\urlprefix }}%
\providecommand \urlprefix  [0]{URL }%
\providecommand \Eprint [0]{\href }%
\providecommand \doibase [0]{http://dx.doi.org/}%
\providecommand \selectlanguage [0]{\@gobble}%
\providecommand \bibinfo  [0]{\@secondoftwo}%
\providecommand \bibfield  [0]{\@secondoftwo}%
\providecommand \translation [1]{[#1]}%
\providecommand \BibitemOpen [0]{}%
\providecommand \bibitemStop [0]{}%
\providecommand \bibitemNoStop [0]{.\EOS\space}%
\providecommand \EOS [0]{\spacefactor3000\relax}%
\providecommand \BibitemShut  [1]{\csname bibitem#1\endcsname}%
\let\auto@bib@innerbib\@empty
\bibitem [{\citenamefont
  {{Peebles}}(1984)}]{Peebles:1984-Tests-of-cosmological-models}%
  \BibitemOpen
  \bibfield  {author} {\bibinfo {author} {\bibfnamefont {P.~J.~E.}\
  \bibnamefont {{Peebles}}},\ }\href {\doibase 10.1086/162425} {\bibfield
  {journal} {\bibinfo  {journal} {\apj}\ }\textbf {\bibinfo {volume} {284}},\
  \bibinfo {pages} {439} (\bibinfo {year} {1984})}\BibitemShut {NoStop}%
\bibitem [{\citenamefont {{Spergel}}\ \emph {et~al.}(2003)\citenamefont
  {{Spergel}}, \citenamefont {{Verde}}, \citenamefont {{Peiris}}, \citenamefont
  {{Komatsu}}, \citenamefont {{Nolta}}, \citenamefont {{Bennett}},
  \citenamefont {{Halpern}}, \citenamefont {{Hinshaw}}, \citenamefont
  {{Jarosik}}, \citenamefont {{Kogut}}, \citenamefont {{Limon}}, \citenamefont
  {{Meyer}}, \citenamefont {{Page}}, \citenamefont {{Tucker}}, \citenamefont
  {{Weiland}}, \citenamefont {{Wollack}},\ and\ \citenamefont
  {{Wright}}}]{Spergel:2003-First-Year-Wilkinson-Microwave-Anisotropy}%
  \BibitemOpen
  \bibfield  {author} {\bibinfo {author} {\bibfnamefont {D.~N.}\ \bibnamefont
  {{Spergel}}}, \bibinfo {author} {\bibfnamefont {L.}~\bibnamefont {{Verde}}},
  \bibinfo {author} {\bibfnamefont {H.~V.}\ \bibnamefont {{Peiris}}}, \bibinfo
  {author} {\bibfnamefont {E.}~\bibnamefont {{Komatsu}}}, \bibinfo {author}
  {\bibfnamefont {M.~R.}\ \bibnamefont {{Nolta}}}, \bibinfo {author}
  {\bibfnamefont {C.~L.}\ \bibnamefont {{Bennett}}}, \bibinfo {author}
  {\bibfnamefont {M.}~\bibnamefont {{Halpern}}}, \bibinfo {author}
  {\bibfnamefont {G.}~\bibnamefont {{Hinshaw}}}, \bibinfo {author}
  {\bibfnamefont {N.}~\bibnamefont {{Jarosik}}}, \bibinfo {author}
  {\bibfnamefont {A.}~\bibnamefont {{Kogut}}}, \bibinfo {author} {\bibfnamefont
  {M.}~\bibnamefont {{Limon}}}, \bibinfo {author} {\bibfnamefont {S.~S.}\
  \bibnamefont {{Meyer}}}, \bibinfo {author} {\bibfnamefont {L.}~\bibnamefont
  {{Page}}}, \bibinfo {author} {\bibfnamefont {G.~S.}\ \bibnamefont
  {{Tucker}}}, \bibinfo {author} {\bibfnamefont {J.~L.}\ \bibnamefont
  {{Weiland}}}, \bibinfo {author} {\bibfnamefont {E.}~\bibnamefont
  {{Wollack}}}, \ and\ \bibinfo {author} {\bibfnamefont {E.~L.}\ \bibnamefont
  {{Wright}}},\ }\href {\doibase 10.1086/377226} {\bibfield  {journal}
  {\bibinfo  {journal} {\apjs}\ }\textbf {\bibinfo {volume} {148}},\ \bibinfo
  {pages} {175} (\bibinfo {year} {2003})},\ \Eprint
  {http://arxiv.org/abs/astro-ph/0302209} {arXiv:astro-ph/0302209 [astro-ph]}
  \BibitemShut {NoStop}%
\bibitem [{\citenamefont {{Komatsu}}\ \emph {et~al.}(2011)\citenamefont
  {{Komatsu}}, \citenamefont {{Smith}}, \citenamefont {{Dunkley}},
  \citenamefont {{Bennett}}, \citenamefont {{Gold}}, \citenamefont {{Hinshaw}},
  \citenamefont {{Jarosik}}, \citenamefont {{Larson}}, \citenamefont {{Nolta}},
  \citenamefont {{Page}}, \citenamefont {{Spergel}}, \citenamefont {{Halpern}},
  \citenamefont {{Hill}}, \citenamefont {{Kogut}}, \citenamefont {{Limon}},
  \citenamefont {{Meyer}}, \citenamefont {{Odegard}}, \citenamefont {{Tucker}},
  \citenamefont {{Weiland}}, \citenamefont {{Wollack}},\ and\ \citenamefont
  {{Wright}}}]{Komatsu:Seven-year-Wilkinson-Microwave-Anisotropy-Probe}%
  \BibitemOpen
  \bibfield  {author} {\bibinfo {author} {\bibfnamefont {E.}~\bibnamefont
  {{Komatsu}}}, \bibinfo {author} {\bibfnamefont {K.~M.}\ \bibnamefont
  {{Smith}}}, \bibinfo {author} {\bibfnamefont {J.}~\bibnamefont {{Dunkley}}},
  \bibinfo {author} {\bibfnamefont {C.~L.}\ \bibnamefont {{Bennett}}}, \bibinfo
  {author} {\bibfnamefont {B.}~\bibnamefont {{Gold}}}, \bibinfo {author}
  {\bibfnamefont {G.}~\bibnamefont {{Hinshaw}}}, \bibinfo {author}
  {\bibfnamefont {N.}~\bibnamefont {{Jarosik}}}, \bibinfo {author}
  {\bibfnamefont {D.}~\bibnamefont {{Larson}}}, \bibinfo {author}
  {\bibfnamefont {M.~R.}\ \bibnamefont {{Nolta}}}, \bibinfo {author}
  {\bibfnamefont {L.}~\bibnamefont {{Page}}}, \bibinfo {author} {\bibfnamefont
  {D.~N.}\ \bibnamefont {{Spergel}}}, \bibinfo {author} {\bibfnamefont
  {M.}~\bibnamefont {{Halpern}}}, \bibinfo {author} {\bibfnamefont {R.~S.}\
  \bibnamefont {{Hill}}}, \bibinfo {author} {\bibfnamefont {A.}~\bibnamefont
  {{Kogut}}}, \bibinfo {author} {\bibfnamefont {M.}~\bibnamefont {{Limon}}},
  \bibinfo {author} {\bibfnamefont {S.~S.}\ \bibnamefont {{Meyer}}}, \bibinfo
  {author} {\bibfnamefont {N.}~\bibnamefont {{Odegard}}}, \bibinfo {author}
  {\bibfnamefont {G.~S.}\ \bibnamefont {{Tucker}}}, \bibinfo {author}
  {\bibfnamefont {J.~L.}\ \bibnamefont {{Weiland}}}, \bibinfo {author}
  {\bibfnamefont {E.}~\bibnamefont {{Wollack}}}, \ and\ \bibinfo {author}
  {\bibfnamefont {E.~L.}\ \bibnamefont {{Wright}}},\ }\href {\doibase
  10.1088/0067-0049/192/2/18} {\bibfield  {journal} {\bibinfo  {journal}
  {\apjs}\ }\textbf {\bibinfo {volume} {192}},\ \bibinfo {eid} {18} (\bibinfo
  {year} {2011})},\ \Eprint {http://arxiv.org/abs/1001.4538} {arXiv:1001.4538
  [astro-ph.CO]} \BibitemShut {NoStop}%
\bibitem [{\citenamefont {{Frenk}}\ and\ \citenamefont
  {{White}}(2012)}]{Frenk:2012-Dark-matter-and-cosmic-structure}%
  \BibitemOpen
  \bibfield  {author} {\bibinfo {author} {\bibfnamefont {C.~S.}\ \bibnamefont
  {{Frenk}}}\ and\ \bibinfo {author} {\bibfnamefont {S.~D.~M.}\ \bibnamefont
  {{White}}},\ }\href {\doibase 10.1002/andp.201200212} {\bibfield  {journal}
  {\bibinfo  {journal} {Annalen der Physik}\ }\textbf {\bibinfo {volume}
  {524}},\ \bibinfo {pages} {507} (\bibinfo {year} {2012})},\ \Eprint
  {http://arxiv.org/abs/1210.0544} {arXiv:1210.0544 [astro-ph.CO]} \BibitemShut
  {NoStop}%
\bibitem [{\citenamefont {Perivolaropoulos}\ and\ \citenamefont
  {Skara}(2022)}]{Perivolaropoulos:2022-Challenges-for}%
  \BibitemOpen
  \bibfield  {author} {\bibinfo {author} {\bibfnamefont {L.}~\bibnamefont
  {Perivolaropoulos}}\ and\ \bibinfo {author} {\bibfnamefont {F.}~\bibnamefont
  {Skara}},\ }\href {\doibase 10.1016/j.newar.2022.101659} {\bibfield
  {journal} {\bibinfo  {journal} {New Astronomy Reviews}\ }\textbf {\bibinfo
  {volume} {95}},\ \bibinfo {pages} {101659} (\bibinfo {year}
  {2022})}\BibitemShut {NoStop}%
\bibitem [{\citenamefont {{Bullock}}\ and\ \citenamefont
  {{Boylan-Kolchin}}(2017)}]{Bullock:2017-Small-Scale-Challenges-to-the}%
  \BibitemOpen
  \bibfield  {author} {\bibinfo {author} {\bibfnamefont {J.~S.}\ \bibnamefont
  {{Bullock}}}\ and\ \bibinfo {author} {\bibfnamefont {M.}~\bibnamefont
  {{Boylan-Kolchin}}},\ }\href {\doibase 10.1146/annurev-astro-091916-055313}
  {\bibfield  {journal} {\bibinfo  {journal} {\araa}\ }\textbf {\bibinfo
  {volume} {55}},\ \bibinfo {pages} {343} (\bibinfo {year} {2017})},\ \Eprint
  {http://arxiv.org/abs/1707.04256} {arXiv:1707.04256 [astro-ph.CO]}
  \BibitemShut {NoStop}%
\bibitem [{\citenamefont {{Flores}}\ and\ \citenamefont
  {{Primack}}(1994)}]{Flores:1994-Observational-and-Theoretical-Constraints}%
  \BibitemOpen
  \bibfield  {author} {\bibinfo {author} {\bibfnamefont {R.~A.}\ \bibnamefont
  {{Flores}}}\ and\ \bibinfo {author} {\bibfnamefont {J.~R.}\ \bibnamefont
  {{Primack}}},\ }\href {\doibase 10.1086/187350} {\bibfield  {journal}
  {\bibinfo  {journal} {\apjl}\ }\textbf {\bibinfo {volume} {427}},\ \bibinfo
  {pages} {L1} (\bibinfo {year} {1994})},\ \Eprint
  {http://arxiv.org/abs/astro-ph/9402004} {arXiv:astro-ph/9402004 [astro-ph]}
  \BibitemShut {NoStop}%
\bibitem [{\citenamefont {de~Blok}(2010)}]{deBlok:2009-The-Core-Cusp-Problem}%
  \BibitemOpen
  \bibfield  {author} {\bibinfo {author} {\bibfnamefont {W.~J.~G.}\
  \bibnamefont {de~Blok}},\ }\href {\doibase 10.1155/2010/789293} {\bibfield
  {journal} {\bibinfo  {journal} {Adv. Astron.}\ }\textbf {\bibinfo {volume}
  {2010}},\ \bibinfo {pages} {789293} (\bibinfo {year} {2010})},\ \Eprint
  {http://arxiv.org/abs/0910.3538} {arXiv:0910.3538 [astro-ph.CO]} \BibitemShut
  {NoStop}%
\bibitem [{\citenamefont {Klypin}\ \emph {et~al.}(1999)\citenamefont {Klypin},
  \citenamefont {Kravtsov}, \citenamefont {Valenzuela},\ and\ \citenamefont
  {Prada}}]{Klypin:1999-Where-Are-the-Missing-Galactic}%
  \BibitemOpen
  \bibfield  {author} {\bibinfo {author} {\bibfnamefont {A.}~\bibnamefont
  {Klypin}}, \bibinfo {author} {\bibfnamefont {A.~V.}\ \bibnamefont
  {Kravtsov}}, \bibinfo {author} {\bibfnamefont {O.}~\bibnamefont
  {Valenzuela}}, \ and\ \bibinfo {author} {\bibfnamefont {F.}~\bibnamefont
  {Prada}},\ }\href {\doibase 10.1086/307643} {\bibfield  {journal} {\bibinfo
  {journal} {The Astrophysical Journal}\ }\textbf {\bibinfo {volume} {522}},\
  \bibinfo {pages} {82} (\bibinfo {year} {1999})}\BibitemShut {NoStop}%
\bibitem [{\citenamefont {Moore}\ \emph {et~al.}(1999)\citenamefont {Moore},
  \citenamefont {Ghigna}, \citenamefont {Governato}, \citenamefont {Lake},
  \citenamefont {Quinn}, \citenamefont {Stadel},\ and\ \citenamefont
  {Tozzi}}]{Moore:1999-Dark-Matter-Substructure}%
  \BibitemOpen
  \bibfield  {author} {\bibinfo {author} {\bibfnamefont {B.}~\bibnamefont
  {Moore}}, \bibinfo {author} {\bibfnamefont {S.}~\bibnamefont {Ghigna}},
  \bibinfo {author} {\bibfnamefont {F.}~\bibnamefont {Governato}}, \bibinfo
  {author} {\bibfnamefont {G.}~\bibnamefont {Lake}}, \bibinfo {author}
  {\bibfnamefont {T.}~\bibnamefont {Quinn}}, \bibinfo {author} {\bibfnamefont
  {J.}~\bibnamefont {Stadel}}, \ and\ \bibinfo {author} {\bibfnamefont
  {P.}~\bibnamefont {Tozzi}},\ }\href {\doibase 10.1086/312287} {\bibfield
  {journal} {\bibinfo  {journal} {The Astrophysical Journal}\ }\textbf
  {\bibinfo {volume} {524}},\ \bibinfo {pages} {L19} (\bibinfo {year}
  {1999})}\BibitemShut {NoStop}%
\bibitem [{\citenamefont {Boylan-Kolchin}\ \emph {et~al.}(2011)\citenamefont
  {Boylan-Kolchin}, \citenamefont {Bullock},\ and\ \citenamefont
  {Kaplinghat}}]{Boylan_Kolchin:2011-Too-big-to-fail}%
  \BibitemOpen
  \bibfield  {author} {\bibinfo {author} {\bibfnamefont {M.}~\bibnamefont
  {Boylan-Kolchin}}, \bibinfo {author} {\bibfnamefont {J.~S.}\ \bibnamefont
  {Bullock}}, \ and\ \bibinfo {author} {\bibfnamefont {M.}~\bibnamefont
  {Kaplinghat}},\ }\href {\doibase 10.1111/j.1745-3933.2011.01074.x} {\bibfield
   {journal} {\bibinfo  {journal} {Monthly Notices of the Royal Astronomical
  Society: Letters}\ }\textbf {\bibinfo {volume} {415}},\ \bibinfo {pages}
  {L40} (\bibinfo {year} {2011})}\BibitemShut {NoStop}%
\bibitem [{\citenamefont {Boylan-Kolchin}\ \emph {et~al.}(2012)\citenamefont
  {Boylan-Kolchin}, \citenamefont {Bullock},\ and\ \citenamefont
  {Kaplinghat}}]{Boylan-Kolchin:2012-The-Milky-Ways-bright-satellites}%
  \BibitemOpen
  \bibfield  {author} {\bibinfo {author} {\bibfnamefont {M.}~\bibnamefont
  {Boylan-Kolchin}}, \bibinfo {author} {\bibfnamefont {J.~S.}\ \bibnamefont
  {Bullock}}, \ and\ \bibinfo {author} {\bibfnamefont {M.}~\bibnamefont
  {Kaplinghat}},\ }\href {\doibase 10.1111/j.1365-2966.2012.20695.x} {\bibfield
   {journal} {\bibinfo  {journal} {Monthly Notices of the Royal Astronomical
  Society}\ }\textbf {\bibinfo {volume} {422}},\ \bibinfo {pages} {1203}
  (\bibinfo {year} {2012})},\ \Eprint
  {http://arxiv.org/abs/https://academic.oup.com/mnras/article-pdf/422/2/1203/3464467/mnras0422-1203.pdf}
  {https://academic.oup.com/mnras/article-pdf/422/2/1203/3464467/mnras0422-1203.pdf}
  \BibitemShut {NoStop}%
\bibitem [{\citenamefont
  {Milgrom}(1983)}]{Milgrom:1983-A-Modification-of-the-Newtonia}%
  \BibitemOpen
  \bibfield  {author} {\bibinfo {author} {\bibfnamefont {M.}~\bibnamefont
  {Milgrom}},\ }\href {\doibase 10.1086/161130} {\bibfield  {journal} {\bibinfo
   {journal} {Astrophysical Journal}\ }\textbf {\bibinfo {volume} {270}},\
  \bibinfo {pages} {365} (\bibinfo {year} {1983})}\BibitemShut {NoStop}%
\bibitem [{\citenamefont {McGaugh}\ \emph {et~al.}(2000)\citenamefont
  {McGaugh}, \citenamefont {Schombert}, \citenamefont {Bothun},\ and\
  \citenamefont {de~Blok}}]{McGaugh:2000-The-baryonic-Tully-Fisher-rela}%
  \BibitemOpen
  \bibfield  {author} {\bibinfo {author} {\bibfnamefont {S.~S.}\ \bibnamefont
  {McGaugh}}, \bibinfo {author} {\bibfnamefont {J.~M.}\ \bibnamefont
  {Schombert}}, \bibinfo {author} {\bibfnamefont {G.~D.}\ \bibnamefont
  {Bothun}}, \ and\ \bibinfo {author} {\bibfnamefont {W.~J.~G.}\ \bibnamefont
  {de~Blok}},\ }\href {\doibase 10.1086/312628} {\bibfield  {journal} {\bibinfo
   {journal} {Astrophysical Journal}\ }\textbf {\bibinfo {volume} {533}},\
  \bibinfo {pages} {L99} (\bibinfo {year} {2000})}\BibitemShut {NoStop}%
\bibitem [{\citenamefont {{Famaey}}\ and\ \citenamefont
  {{McGaugh}}(2013)}]{Famaey:2013-Challenges-fo-CDM-and-MOND}%
  \BibitemOpen
  \bibfield  {author} {\bibinfo {author} {\bibfnamefont {B.}~\bibnamefont
  {{Famaey}}}\ and\ \bibinfo {author} {\bibfnamefont {S.}~\bibnamefont
  {{McGaugh}}},\ }in\ \href {\doibase 10.1088/1742-6596/437/1/012001} {\emph
  {\bibinfo {booktitle} {Journal of Physics Conference Series}}},\ \bibinfo
  {series} {Journal of Physics Conference Series}, Vol.\ \bibinfo {volume}
  {437}\ (\bibinfo {year} {2013})\ p.\ \bibinfo {pages} {012001},\ \Eprint
  {http://arxiv.org/abs/1301.0623} {arXiv:1301.0623 [astro-ph.CO]} \BibitemShut
  {NoStop}%
\bibitem [{\citenamefont {Garrison-Kimmel}\ \emph {et~al.}(2014)\citenamefont
  {Garrison-Kimmel}, \citenamefont {Boylan-Kolchin}, \citenamefont {Bullock},\
  and\ \citenamefont
  {Kirby}}]{Garrison-Kimmel:2014-Too-big-to-fail-in-the-Local-Group}%
  \BibitemOpen
  \bibfield  {author} {\bibinfo {author} {\bibfnamefont {S.}~\bibnamefont
  {Garrison-Kimmel}}, \bibinfo {author} {\bibfnamefont {M.}~\bibnamefont
  {Boylan-Kolchin}}, \bibinfo {author} {\bibfnamefont {J.~S.}\ \bibnamefont
  {Bullock}}, \ and\ \bibinfo {author} {\bibfnamefont {E.~N.}\ \bibnamefont
  {Kirby}},\ }\href {\doibase 10.1093/mnras/stu1477} {\bibfield  {journal}
  {\bibinfo  {journal} {Monthly Notices of the Royal Astronomical Society}\
  }\textbf {\bibinfo {volume} {444}},\ \bibinfo {pages} {222} (\bibinfo {year}
  {2014})},\ \Eprint
  {http://arxiv.org/abs/https://academic.oup.com/mnras/article-pdf/444/1/222/18506913/stu1477.pdf}
  {https://academic.oup.com/mnras/article-pdf/444/1/222/18506913/stu1477.pdf}
  \BibitemShut {NoStop}%
\bibitem [{\citenamefont {{Del Popolo}}\ \emph {et~al.}(2014)\citenamefont
  {{Del Popolo}}, \citenamefont {{Lima}}, \citenamefont {{Fabris}},\ and\
  \citenamefont
  {{Rodrigues}}}]{Popolo:2014-A-unified-solution-to-the-small-scale}%
  \BibitemOpen
  \bibfield  {author} {\bibinfo {author} {\bibfnamefont {A.}~\bibnamefont {{Del
  Popolo}}}, \bibinfo {author} {\bibfnamefont {J.~A.~S.}\ \bibnamefont
  {{Lima}}}, \bibinfo {author} {\bibfnamefont {J.~C.}\ \bibnamefont
  {{Fabris}}}, \ and\ \bibinfo {author} {\bibfnamefont {D.~C.}\ \bibnamefont
  {{Rodrigues}}},\ }\href {\doibase 10.1088/1475-7516/2014/04/021} {\bibfield
  {journal} {\bibinfo  {journal} {\jcap}\ }\textbf {\bibinfo {volume} {2014}},\
  \bibinfo {eid} {021} (\bibinfo {year} {2014})},\ \Eprint
  {http://arxiv.org/abs/1404.3674} {arXiv:1404.3674 [astro-ph.CO]} \BibitemShut
  {NoStop}%
\bibitem [{\citenamefont {{de Blok}}\ and\ \citenamefont
  {{Bosma}}(2002)}]{Blok:2002GALAXIES:-STRUCTURE-GALAXIES}%
  \BibitemOpen
  \bibfield  {author} {\bibinfo {author} {\bibfnamefont {W.~J.~G.}\
  \bibnamefont {{de Blok}}}\ and\ \bibinfo {author} {\bibfnamefont
  {A.}~\bibnamefont {{Bosma}}},\ }\href {\doibase 10.1051/0004-6361:20020080}
  {\bibfield  {journal} {\bibinfo  {journal} {\aap}\ }\textbf {\bibinfo
  {volume} {385}},\ \bibinfo {pages} {816} (\bibinfo {year} {2002})},\ \Eprint
  {http://arxiv.org/abs/astro-ph/0201276} {arXiv:astro-ph/0201276 [astro-ph]}
  \BibitemShut {NoStop}%
\bibitem [{\citenamefont {de~Blok}\ \emph {et~al.}(2003)\citenamefont
  {de~Blok}, \citenamefont {Bosma},\ and\ \citenamefont
  {McGaugh}}]{Blok:2003-Simulating-observations-of-dark-matter}%
  \BibitemOpen
  \bibfield  {author} {\bibinfo {author} {\bibfnamefont {W.~J.~G.}\
  \bibnamefont {de~Blok}}, \bibinfo {author} {\bibfnamefont {A.}~\bibnamefont
  {Bosma}}, \ and\ \bibinfo {author} {\bibfnamefont {S.}~\bibnamefont
  {McGaugh}},\ }\href {\doibase 10.1046/j.1365-8711.2003.06330.x} {\bibfield
  {journal} {\bibinfo  {journal} {Monthly Notices of the Royal Astronomical
  Society}\ }\textbf {\bibinfo {volume} {340}},\ \bibinfo {pages} {657}
  (\bibinfo {year} {2003})},\ \Eprint
  {http://arxiv.org/abs/https://academic.oup.com/mnras/article-pdf/340/2/657/18649385/340-2-657.pdf}
  {https://academic.oup.com/mnras/article-pdf/340/2/657/18649385/340-2-657.pdf}
  \BibitemShut {NoStop}%
\bibitem [{\citenamefont {Swaters}\ \emph {et~al.}(2003)\citenamefont
  {Swaters}, \citenamefont {Madore}, \citenamefont {van~den Bosch},\ and\
  \citenamefont
  {Balcells}}]{Swaters:2002-The-Central-mass-distribution-in-dwarf}%
  \BibitemOpen
  \bibfield  {author} {\bibinfo {author} {\bibfnamefont {R.~A.}\ \bibnamefont
  {Swaters}}, \bibinfo {author} {\bibfnamefont {B.~F.}\ \bibnamefont {Madore}},
  \bibinfo {author} {\bibfnamefont {F.~C.}\ \bibnamefont {van~den Bosch}}, \
  and\ \bibinfo {author} {\bibfnamefont {M.}~\bibnamefont {Balcells}},\ }\href
  {\doibase 10.1086/345426} {\bibfield  {journal} {\bibinfo  {journal}
  {Astrophys. J.}\ }\textbf {\bibinfo {volume} {583}},\ \bibinfo {pages} {732}
  (\bibinfo {year} {2003})},\ \Eprint {http://arxiv.org/abs/astro-ph/0210152}
  {arXiv:astro-ph/0210152} \BibitemShut {NoStop}%
\bibitem [{\citenamefont {Kuzio~de Naray}\ and\ \citenamefont
  {Kaufmann}(2011)}]{Naray:2011-Recovering-cores-and-cusps-in-dark-matter}%
  \BibitemOpen
  \bibfield  {author} {\bibinfo {author} {\bibfnamefont {R.}~\bibnamefont
  {Kuzio~de Naray}}\ and\ \bibinfo {author} {\bibfnamefont {T.}~\bibnamefont
  {Kaufmann}},\ }\href {\doibase 10.1111/j.1365-2966.2011.18656.x} {\bibfield
  {journal} {\bibinfo  {journal} {Monthly Notices of the Royal Astronomical
  Society}\ }\textbf {\bibinfo {volume} {414}},\ \bibinfo {pages} {3617}
  (\bibinfo {year} {2011})},\ \Eprint
  {http://arxiv.org/abs/https://academic.oup.com/mnras/article-pdf/414/4/3617/18715139/mnras0414-3617.pdf}
  {https://academic.oup.com/mnras/article-pdf/414/4/3617/18715139/mnras0414-3617.pdf}
  \BibitemShut {NoStop}%
\bibitem [{\citenamefont {Navarro}\ \emph {et~al.}(1997)\citenamefont
  {Navarro}, \citenamefont {Frenk},\ and\ \citenamefont
  {White}}]{Navarro:1997-A-universal-density-profile-fr}%
  \BibitemOpen
  \bibfield  {author} {\bibinfo {author} {\bibfnamefont {J.~F.}\ \bibnamefont
  {Navarro}}, \bibinfo {author} {\bibfnamefont {C.~S.}\ \bibnamefont {Frenk}},
  \ and\ \bibinfo {author} {\bibfnamefont {S.~D.~M.}\ \bibnamefont {White}},\
  }\href {\doibase 10.1086/304888} {\bibfield  {journal} {\bibinfo  {journal}
  {Astrophysical Journal}\ }\textbf {\bibinfo {volume} {490}},\ \bibinfo
  {pages} {493} (\bibinfo {year} {1997})}\BibitemShut {NoStop}%
\bibitem [{\citenamefont {Navarro}\ \emph {et~al.}(2010)\citenamefont
  {Navarro}, \citenamefont {Ludlow}, \citenamefont {Springel}, \citenamefont
  {Wang}, \citenamefont {Vogelsberger}, \citenamefont {White}, \citenamefont
  {Jenkins}, \citenamefont {Frenk},\ and\ \citenamefont
  {Helmi}}]{Navarro:2010-The-diversity-and-similarity-of-simulated}%
  \BibitemOpen
  \bibfield  {author} {\bibinfo {author} {\bibfnamefont {J.~F.}\ \bibnamefont
  {Navarro}}, \bibinfo {author} {\bibfnamefont {A.}~\bibnamefont {Ludlow}},
  \bibinfo {author} {\bibfnamefont {V.}~\bibnamefont {Springel}}, \bibinfo
  {author} {\bibfnamefont {J.}~\bibnamefont {Wang}}, \bibinfo {author}
  {\bibfnamefont {M.}~\bibnamefont {Vogelsberger}}, \bibinfo {author}
  {\bibfnamefont {S.~D.~M.}\ \bibnamefont {White}}, \bibinfo {author}
  {\bibfnamefont {A.}~\bibnamefont {Jenkins}}, \bibinfo {author} {\bibfnamefont
  {C.~S.}\ \bibnamefont {Frenk}}, \ and\ \bibinfo {author} {\bibfnamefont
  {A.}~\bibnamefont {Helmi}},\ }\href {\doibase
  10.1111/j.1365-2966.2009.15878.x} {\bibfield  {journal} {\bibinfo  {journal}
  {Monthly Notices of the Royal Astronomical Society}\ }\textbf {\bibinfo
  {volume} {402}},\ \bibinfo {pages} {21} (\bibinfo {year} {2010})},\ \Eprint
  {http://arxiv.org/abs/https://academic.oup.com/mnras/article-pdf/402/1/21/18573804/mnras0402-0021.pdf}
  {https://academic.oup.com/mnras/article-pdf/402/1/21/18573804/mnras0402-0021.pdf}
  \BibitemShut {NoStop}%
\bibitem [{\citenamefont {{Diemand}}\ and\ \citenamefont
  {{Moore}}(2011)}]{Diemand:2011-The-Structure-and-Evolution-of-Cold-Dark}%
  \BibitemOpen
  \bibfield  {author} {\bibinfo {author} {\bibfnamefont {J.}~\bibnamefont
  {{Diemand}}}\ and\ \bibinfo {author} {\bibfnamefont {B.}~\bibnamefont
  {{Moore}}},\ }\href {\doibase 10.1166/asl.2011.1211} {\bibfield  {journal}
  {\bibinfo  {journal} {Advanced Science Letters}\ }\textbf {\bibinfo {volume}
  {4}},\ \bibinfo {pages} {297} (\bibinfo {year} {2011})},\ \Eprint
  {http://arxiv.org/abs/0906.4340} {arXiv:0906.4340 [astro-ph.CO]} \BibitemShut
  {NoStop}%
\bibitem [{\citenamefont {Governato}\ \emph {et~al.}(2010)\citenamefont
  {Governato}, \citenamefont {Brook}, \citenamefont {Mayer}, \citenamefont
  {Brooks}, \citenamefont {Rhee}, \citenamefont {Wadsley}, \citenamefont
  {Jonsson}, \citenamefont {Willman}, \citenamefont {Stinson}, \citenamefont
  {Quinn},\ and\ \citenamefont
  {Madau}}]{Governato:2010-Bulgeless-dwarf-galaxies-and-dark-matter-cores}%
  \BibitemOpen
  \bibfield  {author} {\bibinfo {author} {\bibfnamefont {F.}~\bibnamefont
  {Governato}}, \bibinfo {author} {\bibfnamefont {C.}~\bibnamefont {Brook}},
  \bibinfo {author} {\bibfnamefont {L.}~\bibnamefont {Mayer}}, \bibinfo
  {author} {\bibfnamefont {A.}~\bibnamefont {Brooks}}, \bibinfo {author}
  {\bibfnamefont {G.}~\bibnamefont {Rhee}}, \bibinfo {author} {\bibfnamefont
  {J.}~\bibnamefont {Wadsley}}, \bibinfo {author} {\bibfnamefont
  {P.}~\bibnamefont {Jonsson}}, \bibinfo {author} {\bibfnamefont
  {B.}~\bibnamefont {Willman}}, \bibinfo {author} {\bibfnamefont
  {G.}~\bibnamefont {Stinson}}, \bibinfo {author} {\bibfnamefont
  {T.}~\bibnamefont {Quinn}}, \ and\ \bibinfo {author} {\bibfnamefont
  {P.}~\bibnamefont {Madau}},\ }\href {\doibase 10.1038/nature08640} {\bibfield
   {journal} {\bibinfo  {journal} {Nature}\ }\textbf {\bibinfo {volume}
  {463}},\ \bibinfo {pages} {203} (\bibinfo {year} {2010})}\BibitemShut
  {NoStop}%
\bibitem [{\citenamefont {{McKeown}}\ \emph {et~al.}(2022)\citenamefont
  {{McKeown}}, \citenamefont {{Bullock}}, \citenamefont {{Mercado}},
  \citenamefont {{Hafen}}, \citenamefont {{Boylan-Kolchin}}, \citenamefont
  {{Wetzel}}, \citenamefont {{Necib}}, \citenamefont {{Hopkins}},\ and\
  \citenamefont
  {{Yu}}}]{McKeown:2022-Amplified-J-factors-in-the-Galactic-Centre}%
  \BibitemOpen
  \bibfield  {author} {\bibinfo {author} {\bibfnamefont {D.}~\bibnamefont
  {{McKeown}}}, \bibinfo {author} {\bibfnamefont {J.~S.}\ \bibnamefont
  {{Bullock}}}, \bibinfo {author} {\bibfnamefont {F.~J.}\ \bibnamefont
  {{Mercado}}}, \bibinfo {author} {\bibfnamefont {Z.}~\bibnamefont {{Hafen}}},
  \bibinfo {author} {\bibfnamefont {M.}~\bibnamefont {{Boylan-Kolchin}}},
  \bibinfo {author} {\bibfnamefont {A.}~\bibnamefont {{Wetzel}}}, \bibinfo
  {author} {\bibfnamefont {L.}~\bibnamefont {{Necib}}}, \bibinfo {author}
  {\bibfnamefont {P.~F.}\ \bibnamefont {{Hopkins}}}, \ and\ \bibinfo {author}
  {\bibfnamefont {S.}~\bibnamefont {{Yu}}},\ }\href {\doibase
  10.1093/mnras/stac966} {\bibfield  {journal} {\bibinfo  {journal} {\mnras}\
  }\textbf {\bibinfo {volume} {513}},\ \bibinfo {pages} {55} (\bibinfo {year}
  {2022})},\ \Eprint {http://arxiv.org/abs/2111.03076} {arXiv:2111.03076
  [astro-ph.GA]} \BibitemShut {NoStop}%
\bibitem [{\citenamefont {Del~Popolo}\ and\ \citenamefont
  {Le~Delliou}(2017)}]{DelPopolo:2017-Small-scale-problems-of-the}%
  \BibitemOpen
  \bibfield  {author} {\bibinfo {author} {\bibfnamefont {A.}~\bibnamefont
  {Del~Popolo}}\ and\ \bibinfo {author} {\bibfnamefont {M.}~\bibnamefont
  {Le~Delliou}},\ }\href {\doibase 10.3390/galaxies5010017} {\bibfield
  {journal} {\bibinfo  {journal} {Galaxies}\ }\textbf {\bibinfo {volume} {5}},\
  \bibinfo {pages} {17} (\bibinfo {year} {2017})},\ \Eprint
  {http://arxiv.org/abs/1606.07790} {arXiv:1606.07790 [astro-ph.CO]}
  \BibitemShut {NoStop}%
\bibitem [{\citenamefont {{Navarro}}\ \emph {et~al.}(1996)\citenamefont
  {{Navarro}}, \citenamefont {{Eke}},\ and\ \citenamefont
  {{Frenk}}}]{Navarro:1996-The-cores-of-dwarf-galaxy-haloes}%
  \BibitemOpen
  \bibfield  {author} {\bibinfo {author} {\bibfnamefont {J.~F.}\ \bibnamefont
  {{Navarro}}}, \bibinfo {author} {\bibfnamefont {V.~R.}\ \bibnamefont
  {{Eke}}}, \ and\ \bibinfo {author} {\bibfnamefont {C.~S.}\ \bibnamefont
  {{Frenk}}},\ }\href {\doibase 10.1093/mnras/283.3.L72} {\bibfield  {journal}
  {\bibinfo  {journal} {\mnras}\ }\textbf {\bibinfo {volume} {283}},\ \bibinfo
  {pages} {L72} (\bibinfo {year} {1996})},\ \Eprint
  {http://arxiv.org/abs/astro-ph/9610187} {arXiv:astro-ph/9610187 [astro-ph]}
  \BibitemShut {NoStop}%
\bibitem [{\citenamefont {{Oh}}\ \emph {et~al.}(2011)\citenamefont {{Oh}},
  \citenamefont {{Brook}}, \citenamefont {{Governato}}, \citenamefont
  {{Brinks}}, \citenamefont {{Mayer}}, \citenamefont {{de Blok}}, \citenamefont
  {{Brooks}},\ and\ \citenamefont
  {{Walter}}}]{Oh:2011-The-Central-Slope-of-Dark-Matter-Cores}%
  \BibitemOpen
  \bibfield  {author} {\bibinfo {author} {\bibfnamefont {S.-H.}\ \bibnamefont
  {{Oh}}}, \bibinfo {author} {\bibfnamefont {C.}~\bibnamefont {{Brook}}},
  \bibinfo {author} {\bibfnamefont {F.}~\bibnamefont {{Governato}}}, \bibinfo
  {author} {\bibfnamefont {E.}~\bibnamefont {{Brinks}}}, \bibinfo {author}
  {\bibfnamefont {L.}~\bibnamefont {{Mayer}}}, \bibinfo {author} {\bibfnamefont
  {W.~J.~G.}\ \bibnamefont {{de Blok}}}, \bibinfo {author} {\bibfnamefont
  {A.}~\bibnamefont {{Brooks}}}, \ and\ \bibinfo {author} {\bibfnamefont
  {F.}~\bibnamefont {{Walter}}},\ }\href {\doibase 10.1088/0004-6256/142/1/24}
  {\bibfield  {journal} {\bibinfo  {journal} {\aj}\ }\textbf {\bibinfo {volume}
  {142}},\ \bibinfo {eid} {24} (\bibinfo {year} {2011})},\ \Eprint
  {http://arxiv.org/abs/1011.2777} {arXiv:1011.2777 [astro-ph.CO]} \BibitemShut
  {NoStop}%
\bibitem [{\citenamefont {Benítez-Llambay}\ \emph {et~al.}(2019)\citenamefont
  {Benítez-Llambay}, \citenamefont {Frenk}, \citenamefont {Ludlow},\ and\
  \citenamefont {Navarro}}]{Benitez:2019-Baryon-induced-dark-matter-cores}%
  \BibitemOpen
  \bibfield  {author} {\bibinfo {author} {\bibfnamefont {A.}~\bibnamefont
  {Benítez-Llambay}}, \bibinfo {author} {\bibfnamefont {C.~S.}\ \bibnamefont
  {Frenk}}, \bibinfo {author} {\bibfnamefont {A.~D.}\ \bibnamefont {Ludlow}}, \
  and\ \bibinfo {author} {\bibfnamefont {J.~F.}\ \bibnamefont {Navarro}},\
  }\href {\doibase 10.1093/mnras/stz1890} {\bibfield  {journal} {\bibinfo
  {journal} {Monthly Notices of the Royal Astronomical Society}\ }\textbf
  {\bibinfo {volume} {488}},\ \bibinfo {pages} {2387} (\bibinfo {year}
  {2019})},\ \Eprint
  {http://arxiv.org/abs/https://academic.oup.com/mnras/article-pdf/488/2/2387/28979701/stz1890.pdf}
  {https://academic.oup.com/mnras/article-pdf/488/2/2387/28979701/stz1890.pdf}
  \BibitemShut {NoStop}%
\bibitem [{\citenamefont {Spergel}\ and\ \citenamefont
  {Steinhardt}(2000)}]{Spergel:2000-Observational-Evidence-for-Self-Interacting-Cold-Dark-Matter}%
  \BibitemOpen
  \bibfield  {author} {\bibinfo {author} {\bibfnamefont {D.~N.}\ \bibnamefont
  {Spergel}}\ and\ \bibinfo {author} {\bibfnamefont {P.~J.}\ \bibnamefont
  {Steinhardt}},\ }\href {\doibase 10.1103/PhysRevLett.84.3760} {\bibfield
  {journal} {\bibinfo  {journal} {Phys. Rev. Lett.}\ }\textbf {\bibinfo
  {volume} {84}},\ \bibinfo {pages} {3760} (\bibinfo {year}
  {2000})}\BibitemShut {NoStop}%
\bibitem [{\citenamefont {Rocha}\ \emph {et~al.}(2013)\citenamefont {Rocha},
  \citenamefont {Peter}, \citenamefont {Bullock}, \citenamefont {Kaplinghat},
  \citenamefont {Garrison-Kimmel}, \citenamefont {Oñorbe},\ and\ \citenamefont
  {Moustakas}}]{Rocha:2013-Cosmological-simulations-with-self-interacting-dark-matter}%
  \BibitemOpen
  \bibfield  {author} {\bibinfo {author} {\bibfnamefont {M.}~\bibnamefont
  {Rocha}}, \bibinfo {author} {\bibfnamefont {A.~H.~G.}\ \bibnamefont {Peter}},
  \bibinfo {author} {\bibfnamefont {J.~S.}\ \bibnamefont {Bullock}}, \bibinfo
  {author} {\bibfnamefont {M.}~\bibnamefont {Kaplinghat}}, \bibinfo {author}
  {\bibfnamefont {S.}~\bibnamefont {Garrison-Kimmel}}, \bibinfo {author}
  {\bibfnamefont {J.}~\bibnamefont {Oñorbe}}, \ and\ \bibinfo {author}
  {\bibfnamefont {L.~A.}\ \bibnamefont {Moustakas}},\ }\href {\doibase
  10.1093/mnras/sts514} {\bibfield  {journal} {\bibinfo  {journal} {Monthly
  Notices of the Royal Astronomical Society}\ }\textbf {\bibinfo {volume}
  {430}},\ \bibinfo {pages} {81} (\bibinfo {year} {2013})},\ \Eprint
  {http://arxiv.org/abs/https://academic.oup.com/mnras/article-pdf/430/1/81/3064615/sts514.pdf}
  {https://academic.oup.com/mnras/article-pdf/430/1/81/3064615/sts514.pdf}
  \BibitemShut {NoStop}%
\bibitem [{\citenamefont {{Peter}}\ \emph {et~al.}(2013)\citenamefont
  {{Peter}}, \citenamefont {{Rocha}}, \citenamefont {{Bullock}},\ and\
  \citenamefont
  {{Kaplinghat}}}]{Peter:2013-Cosmological-simulations-with-self-interacting-dark-matter}%
  \BibitemOpen
  \bibfield  {author} {\bibinfo {author} {\bibfnamefont {A.~H.~G.}\
  \bibnamefont {{Peter}}}, \bibinfo {author} {\bibfnamefont {M.}~\bibnamefont
  {{Rocha}}}, \bibinfo {author} {\bibfnamefont {J.~S.}\ \bibnamefont
  {{Bullock}}}, \ and\ \bibinfo {author} {\bibfnamefont {M.}~\bibnamefont
  {{Kaplinghat}}},\ }\href {\doibase 10.1093/mnras/sts535} {\bibfield
  {journal} {\bibinfo  {journal} {\mnras}\ }\textbf {\bibinfo {volume} {430}},\
  \bibinfo {pages} {105} (\bibinfo {year} {2013})},\ \Eprint
  {http://arxiv.org/abs/1208.3026} {arXiv:1208.3026 [astro-ph.CO]} \BibitemShut
  {NoStop}%
\bibitem [{\citenamefont {Rubin}\ and\ \citenamefont
  {Ford}(1970)}]{Rubin:1970-Rotation-of-Andromeda-Nebula-f}%
  \BibitemOpen
  \bibfield  {author} {\bibinfo {author} {\bibfnamefont {V.~C.}\ \bibnamefont
  {Rubin}}\ and\ \bibinfo {author} {\bibfnamefont {W.~K.}\ \bibnamefont
  {Ford}},\ }\href {\doibase 10.1086/150317} {\bibfield  {journal} {\bibinfo
  {journal} {Astrophysical Journal}\ }\textbf {\bibinfo {volume} {159}},\
  \bibinfo {pages} {379} (\bibinfo {year} {1970})}\BibitemShut {NoStop}%
\bibitem [{\citenamefont {Rubin}\ \emph {et~al.}(1980)\citenamefont {Rubin},
  \citenamefont {Ford},\ and\ \citenamefont
  {Thonnard}}]{Rubin:1980-Rotational-Properties-of-21-Sc}%
  \BibitemOpen
  \bibfield  {author} {\bibinfo {author} {\bibfnamefont {V.~C.}\ \bibnamefont
  {Rubin}}, \bibinfo {author} {\bibfnamefont {W.~K.}\ \bibnamefont {Ford}}, \
  and\ \bibinfo {author} {\bibfnamefont {N.}~\bibnamefont {Thonnard}},\ }\href
  {\doibase 10.1086/158003} {\bibfield  {journal} {\bibinfo  {journal}
  {Astrophysical Journal}\ }\textbf {\bibinfo {volume} {238}},\ \bibinfo
  {pages} {471} (\bibinfo {year} {1980})}\BibitemShut {NoStop}%
\bibitem [{\citenamefont
  {Taylor}(1935)}]{Taylor:1935-Statistical-theory-of-turbulan}%
  \BibitemOpen
  \bibfield  {author} {\bibinfo {author} {\bibfnamefont {G.~I.}\ \bibnamefont
  {Taylor}},\ }\href {\doibase 10.1098/rspa.1935.0158} {\bibfield  {journal}
  {\bibinfo  {journal} {Proceedings of the royal society A}\ }\textbf {\bibinfo
  {volume} {151}},\ \bibinfo {pages} {421} (\bibinfo {year}
  {1935})}\BibitemShut {NoStop}%
\bibitem [{\citenamefont {de~Karman}\ and\ \citenamefont
  {Howarth}(1938)}]{de_Karman:1938-On-the-statistical-theory-of-i}%
  \BibitemOpen
  \bibfield  {author} {\bibinfo {author} {\bibfnamefont {T.}~\bibnamefont
  {de~Karman}}\ and\ \bibinfo {author} {\bibfnamefont {L.}~\bibnamefont
  {Howarth}},\ }\href {\doibase 10.1098/rspa.1938.0013} {\bibfield  {journal}
  {\bibinfo  {journal} {Proceedings of the Royal Society of London Series
  a-Mathematical and Physical Sciences}\ }\textbf {\bibinfo {volume} {164}},\
  \bibinfo {pages} {0192} (\bibinfo {year} {1938})}\BibitemShut {NoStop}%
\bibitem [{\citenamefont
  {Batchelor}(1953)}]{Batchelor:1953-The-Theory-of-Homogeneous-Turb}%
  \BibitemOpen
  \bibfield  {author} {\bibinfo {author} {\bibfnamefont {G.~K.}\ \bibnamefont
  {Batchelor}},\ }\href@noop {} {\emph {\bibinfo {title} {The Theory of
  Homogeneous Turbulence}}}\ (\bibinfo  {publisher} {Cambridge University
  Press},\ \bibinfo {address} {Cambridge, UK},\ \bibinfo {year}
  {1953})\BibitemShut {NoStop}%
\bibitem [{\citenamefont
  {Richardson}(1922)}]{Richardson:1922-Weather-Prediction-by-Numerica}%
  \BibitemOpen
  \bibfield  {author} {\bibinfo {author} {\bibfnamefont {L.~F.}\ \bibnamefont
  {Richardson}},\ }\href@noop {} {\emph {\bibinfo {title} {Weather Prediction
  by Numerical Process}}}\ (\bibinfo  {publisher} {Cambridge University
  Press},\ \bibinfo {address} {Cambridge, UK},\ \bibinfo {year}
  {1922})\BibitemShut {NoStop}%
\bibitem [{\citenamefont
  {Xu}(2022{\natexlab{a}})}]{Xu:2022-Dark_matter-flow-and-hydrodynamic-turbulence-presentation}%
  \BibitemOpen
  \bibfield  {author} {\bibinfo {author} {\bibfnamefont {Z.}~\bibnamefont
  {Xu}},\ }\href {\doibase 10.5281/zenodo.6569901} {\enquote {\bibinfo {title}
  {A comparative study of dark matter flow \& hydrodynamic turbulence and its
  applications},}\ } (\bibinfo {year} {2022}{\natexlab{a}})\BibitemShut
  {NoStop}%
\bibitem [{\citenamefont
  {Xu}(2021{\natexlab{a}})}]{Xu:2021-Inverse-mass-cascade-mass-function}%
  \BibitemOpen
  \bibfield  {author} {\bibinfo {author} {\bibfnamefont {Z.}~\bibnamefont
  {Xu}},\ }\href {\doibase 10.48550/ARXIV.2109.09985} {\bibfield  {journal}
  {\bibinfo  {journal} {arXiv e-prints}\ ,\ \bibinfo {pages}
  {arXiv:2109.09985}} (\bibinfo {year} {2021}{\natexlab{a}})}\BibitemShut
  {NoStop}%
\bibitem [{\citenamefont
  {Xu}(2021{\natexlab{b}})}]{Xu:2021-Inverse-and-direct-cascade-of-}%
  \BibitemOpen
  \bibfield  {author} {\bibinfo {author} {\bibfnamefont {Z.}~\bibnamefont
  {Xu}},\ }\href {\doibase 10.48550/ARXIV.2110.13885} {\bibfield  {journal}
  {\bibinfo  {journal} {arXiv e-prints}\ ,\ \bibinfo {pages}
  {arXiv:2110.13885}} (\bibinfo {year} {2021}{\natexlab{b}})}\BibitemShut
  {NoStop}%
\bibitem [{\citenamefont
  {Xu}(2023{\natexlab{a}})}]{Xu:2023-On-the-statistical-theory-of-self-gravitating}%
  \BibitemOpen
  \bibfield  {author} {\bibinfo {author} {\bibfnamefont {Z.}~\bibnamefont
  {Xu}},\ }\href {\doibase 10.1063/5.0151129} {\bibfield  {journal} {\bibinfo
  {journal} {Physics of Fluids}\ }\textbf {\bibinfo {volume} {35}},\ \bibinfo
  {pages} {077105} (\bibinfo {year} {2023}{\natexlab{a}})},\ \Eprint
  {http://arxiv.org/abs/2202.00910} {arXiv:2202.00910 [astro-ph]} \BibitemShut
  {NoStop}%
\bibitem [{\citenamefont
  {Xu}(2023{\natexlab{b}})}]{Xu:2023-Maximum-entropy-distributions-of-dark-matter}%
  \BibitemOpen
  \bibfield  {author} {\bibinfo {author} {\bibfnamefont {Z.}~\bibnamefont
  {Xu}},\ }\href {\doibase 10.1051/0004-6361/202346429} {\bibfield  {journal}
  {\bibinfo  {journal} {A\&A}\ }\textbf {\bibinfo {volume} {675}},\ \bibinfo
  {pages} {A92} (\bibinfo {year} {2023}{\natexlab{b}})},\ \Eprint
  {http://arxiv.org/abs/2110.03126} {arXiv:2110.03126 [astro-ph]} \BibitemShut
  {NoStop}%
\bibitem [{\citenamefont {Neyman}\ and\ \citenamefont
  {Scott}(1952)}]{Neyman:1952-A-Theory-of-the-Spatial-Distri}%
  \BibitemOpen
  \bibfield  {author} {\bibinfo {author} {\bibfnamefont {J.}~\bibnamefont
  {Neyman}}\ and\ \bibinfo {author} {\bibfnamefont {E.~L.}\ \bibnamefont
  {Scott}},\ }\href {\doibase 10.1086/145599} {\bibfield  {journal} {\bibinfo
  {journal} {Astrophysical Journal}\ }\textbf {\bibinfo {volume} {116}},\
  \bibinfo {pages} {144} (\bibinfo {year} {1952})}\BibitemShut {NoStop}%
\bibitem [{\citenamefont {Cooray}\ and\ \citenamefont
  {Sheth}(2002)}]{Cooray:2002-Halo-models-of-large-scale-str}%
  \BibitemOpen
  \bibfield  {author} {\bibinfo {author} {\bibfnamefont {A.}~\bibnamefont
  {Cooray}}\ and\ \bibinfo {author} {\bibfnamefont {R.}~\bibnamefont {Sheth}},\
  }\href {\doibase 10.1016/S0370-1573(02)00276-4} {\bibfield  {journal}
  {\bibinfo  {journal} {Physics Reports-Review Section of Physics Letters}\
  }\textbf {\bibinfo {volume} {372}},\ \bibinfo {pages} {1} (\bibinfo {year}
  {2002})}\BibitemShut {NoStop}%
\bibitem [{\citenamefont
  {Xu}(2023{\natexlab{c}})}]{Xu:2023-Dark-matter-halo-mass-functions-and}%
  \BibitemOpen
  \bibfield  {author} {\bibinfo {author} {\bibfnamefont {Z.}~\bibnamefont
  {Xu}},\ }\href {\doibase 10.1038/s41598-023-42958-6} {\bibfield  {journal}
  {\bibinfo  {journal} {Scientific Reports}\ } (\bibinfo {year}
  {2023}{\natexlab{c}}),\ 10.1038/s41598-023-42958-6},\ \Eprint
  {http://arxiv.org/abs/2210.01200} {arXiv:2210.01200 [astro-ph]} \BibitemShut
  {NoStop}%
\bibitem [{\citenamefont
  {Peebles}(1980)}]{Peebles:1980-The-Large-Scale-Structure-of-t}%
  \BibitemOpen
  \bibfield  {author} {\bibinfo {author} {\bibfnamefont {P.~J.~E.}\
  \bibnamefont {Peebles}},\ }\href@noop {} {\emph {\bibinfo {title} {The
  Large-Scale Structure of the Universe}}}\ (\bibinfo  {publisher} {Princeton
  University Press},\ \bibinfo {address} {Princeton, NJ},\ \bibinfo {year}
  {1980})\BibitemShut {NoStop}%
\bibitem [{\citenamefont {Frenk}\ \emph {et~al.}(2000)\citenamefont {Frenk},
  \citenamefont {Colberg}, \citenamefont {Couchman}, \citenamefont
  {Efstathiou}, \citenamefont {Evrard}, \citenamefont {Jenkins}, \citenamefont
  {MacFarland}, \citenamefont {Moore}, \citenamefont {Peacock}, \citenamefont
  {Pearce}, \citenamefont {Thomas}, \citenamefont {White},\ and\ \citenamefont
  {Yoshida.}}]{Frenk:2000-Public-Release-of-N-body-simul}%
  \BibitemOpen
  \bibfield  {author} {\bibinfo {author} {\bibfnamefont {C.~S.}\ \bibnamefont
  {Frenk}}, \bibinfo {author} {\bibfnamefont {J.~M.}\ \bibnamefont {Colberg}},
  \bibinfo {author} {\bibfnamefont {H.~M.~P.}\ \bibnamefont {Couchman}},
  \bibinfo {author} {\bibfnamefont {G.}~\bibnamefont {Efstathiou}}, \bibinfo
  {author} {\bibfnamefont {A.~E.}\ \bibnamefont {Evrard}}, \bibinfo {author}
  {\bibfnamefont {A.}~\bibnamefont {Jenkins}}, \bibinfo {author} {\bibfnamefont
  {T.~J.}\ \bibnamefont {MacFarland}}, \bibinfo {author} {\bibfnamefont
  {B.}~\bibnamefont {Moore}}, \bibinfo {author} {\bibfnamefont {J.~A.}\
  \bibnamefont {Peacock}}, \bibinfo {author} {\bibfnamefont {F.~R.}\
  \bibnamefont {Pearce}}, \bibinfo {author} {\bibfnamefont {P.~A.}\
  \bibnamefont {Thomas}}, \bibinfo {author} {\bibfnamefont {S.~D.~M.}\
  \bibnamefont {White}}, \ and\ \bibinfo {author} {\bibfnamefont
  {N.}~\bibnamefont {Yoshida.}},\ }\href {\doibase
  10.48550/arXiv.astro-ph/0007362} {\bibfield  {journal} {\bibinfo  {journal}
  {arXiv:astro-ph/0007362v1}\ } (\bibinfo {year} {2000}),\
  10.48550/arXiv.astro-ph/0007362}\BibitemShut {NoStop}%
\bibitem [{\citenamefont {Jenkins}\ \emph {et~al.}(1998)\citenamefont
  {Jenkins}, \citenamefont {Frenk}, \citenamefont {Pearce}, \citenamefont
  {Thomas}, \citenamefont {Colberg}, \citenamefont {White}, \citenamefont
  {Couchman}, \citenamefont {Peacock}, \citenamefont {Efstathiou},\ and\
  \citenamefont {Nelson}}]{Jenkins:1998-Evolution-of-structure-in-cold}%
  \BibitemOpen
  \bibfield  {author} {\bibinfo {author} {\bibfnamefont {A.}~\bibnamefont
  {Jenkins}}, \bibinfo {author} {\bibfnamefont {C.~S.}\ \bibnamefont {Frenk}},
  \bibinfo {author} {\bibfnamefont {F.~R.}\ \bibnamefont {Pearce}}, \bibinfo
  {author} {\bibfnamefont {P.~A.}\ \bibnamefont {Thomas}}, \bibinfo {author}
  {\bibfnamefont {J.~M.}\ \bibnamefont {Colberg}}, \bibinfo {author}
  {\bibfnamefont {S.~D.~M.}\ \bibnamefont {White}}, \bibinfo {author}
  {\bibfnamefont {H.~M.~P.}\ \bibnamefont {Couchman}}, \bibinfo {author}
  {\bibfnamefont {J.~A.}\ \bibnamefont {Peacock}}, \bibinfo {author}
  {\bibfnamefont {G.}~\bibnamefont {Efstathiou}}, \ and\ \bibinfo {author}
  {\bibfnamefont {A.~H.}\ \bibnamefont {Nelson}},\ }\href {\doibase
  10.1086/305615} {\bibfield  {journal} {\bibinfo  {journal} {Astrophysical
  Journal}\ }\textbf {\bibinfo {volume} {499}},\ \bibinfo {pages} {20}
  (\bibinfo {year} {1998})}\BibitemShut {NoStop}%
\bibitem [{\citenamefont
  {Irvine}(1961)}]{Irvine:1961-Local-Irregularities-in-a-Univ}%
  \BibitemOpen
  \bibfield  {author} {\bibinfo {author} {\bibfnamefont {W.~M.}\ \bibnamefont
  {Irvine}},\ }\emph {\bibinfo {title} {Local Irregularities in a Universe
  Satisfying the Cosmological Principle}},\ \href@noop {} {\bibinfo {type}
  {Thesis}},\ \bibinfo  {school} {HARVARD UNIVERSITY} (\bibinfo {year}
  {1961})\BibitemShut {NoStop}%
\bibitem [{\citenamefont
  {Layzer}(1963)}]{Layzer:1963-A-Preface-to-Cosmogony--I--The}%
  \BibitemOpen
  \bibfield  {author} {\bibinfo {author} {\bibfnamefont {D.}~\bibnamefont
  {Layzer}},\ }\href {\doibase 10.1086/147625} {\bibfield  {journal} {\bibinfo
  {journal} {Astrophysical Journal}\ }\textbf {\bibinfo {volume} {138}},\
  \bibinfo {pages} {174} (\bibinfo {year} {1963})}\BibitemShut {NoStop}%
\bibitem [{\citenamefont
  {Xu}(2022{\natexlab{b}})}]{Xu:2022-The-evolution-of-energy--momen}%
  \BibitemOpen
  \bibfield  {author} {\bibinfo {author} {\bibfnamefont {Z.}~\bibnamefont
  {Xu}},\ }\href {\doibase 10.48550/ARXIV.2202.04054} {\bibfield  {journal}
  {\bibinfo  {journal} {arXiv e-prints}\ ,\ \bibinfo {pages}
  {arXiv:2202.04054}} (\bibinfo {year} {2022}{\natexlab{b}})}\BibitemShut
  {NoStop}%
\bibitem [{\citenamefont
  {Xu}(2022{\natexlab{c}})}]{Xu:2022-Postulating-dark-matter-partic}%
  \BibitemOpen
  \bibfield  {author} {\bibinfo {author} {\bibfnamefont {Z.}~\bibnamefont
  {Xu}},\ }\href {\doibase 10.48550/ARXIV.2202.07240} {\bibfield  {journal}
  {\bibinfo  {journal} {arXiv e-prints}\ ,\ \bibinfo {pages}
  {arXiv:2202.07240}} (\bibinfo {year} {2022}{\natexlab{c}})}\BibitemShut
  {NoStop}%
\bibitem [{\citenamefont
  {Xu}(2022{\natexlab{d}})}]{Xu:2022-Two-thirds-law-for-pairwise-ve}%
  \BibitemOpen
  \bibfield  {author} {\bibinfo {author} {\bibfnamefont {Z.}~\bibnamefont
  {Xu}},\ }\href {\doibase 10.48550/ARXIV.2202.06515} {\bibfield  {journal}
  {\bibinfo  {journal} {arXiv e-prints}\ ,\ \bibinfo {pages}
  {arXiv:2202.06515}} (\bibinfo {year} {2022}{\natexlab{d}})}\BibitemShut
  {NoStop}%
\bibitem [{\citenamefont {Nelson}\ \emph {et~al.}(2015)\citenamefont {Nelson},
  \citenamefont {Pillepich}, \citenamefont {Genel}, \citenamefont
  {Vogelsberger}, \citenamefont {Springel}, \citenamefont {Torrey},
  \citenamefont {Rodriguez-Gomez}, \citenamefont {Sijacki}, \citenamefont
  {Snyder}, \citenamefont {Griffen}, \citenamefont {Marinacci}, \citenamefont
  {Blecha}, \citenamefont {Sales}, \citenamefont {Xu},\ and\ \citenamefont
  {Hernquist}}]{NELSON:2015-The-illustris-simulation}%
  \BibitemOpen
  \bibfield  {author} {\bibinfo {author} {\bibfnamefont {D.}~\bibnamefont
  {Nelson}}, \bibinfo {author} {\bibfnamefont {A.}~\bibnamefont {Pillepich}},
  \bibinfo {author} {\bibfnamefont {S.}~\bibnamefont {Genel}}, \bibinfo
  {author} {\bibfnamefont {M.}~\bibnamefont {Vogelsberger}}, \bibinfo {author}
  {\bibfnamefont {V.}~\bibnamefont {Springel}}, \bibinfo {author}
  {\bibfnamefont {P.}~\bibnamefont {Torrey}}, \bibinfo {author} {\bibfnamefont
  {V.}~\bibnamefont {Rodriguez-Gomez}}, \bibinfo {author} {\bibfnamefont
  {D.}~\bibnamefont {Sijacki}}, \bibinfo {author} {\bibfnamefont
  {G.}~\bibnamefont {Snyder}}, \bibinfo {author} {\bibfnamefont
  {B.}~\bibnamefont {Griffen}}, \bibinfo {author} {\bibfnamefont
  {F.}~\bibnamefont {Marinacci}}, \bibinfo {author} {\bibfnamefont
  {L.}~\bibnamefont {Blecha}}, \bibinfo {author} {\bibfnamefont
  {L.}~\bibnamefont {Sales}}, \bibinfo {author} {\bibfnamefont
  {D.}~\bibnamefont {Xu}}, \ and\ \bibinfo {author} {\bibfnamefont
  {L.}~\bibnamefont {Hernquist}},\ }\href {\doibase
  https://doi.org/10.1016/j.ascom.2015.09.003} {\bibfield  {journal} {\bibinfo
  {journal} {Astronomy and Computing}\ }\textbf {\bibinfo {volume} {13}},\
  \bibinfo {pages} {12} (\bibinfo {year} {2015})}\BibitemShut {NoStop}%
\bibitem [{\citenamefont {Zhao}\ \emph {et~al.}(2009)\citenamefont {Zhao},
  \citenamefont {Jing}, \citenamefont {Mo},\ and\ \citenamefont
  {Borner}}]{Zhao:2009-Accurate-Universal-Models-for-}%
  \BibitemOpen
  \bibfield  {author} {\bibinfo {author} {\bibfnamefont {D.~H.}\ \bibnamefont
  {Zhao}}, \bibinfo {author} {\bibfnamefont {Y.~P.}\ \bibnamefont {Jing}},
  \bibinfo {author} {\bibfnamefont {H.~J.}\ \bibnamefont {Mo}}, \ and\ \bibinfo
  {author} {\bibfnamefont {G.}~\bibnamefont {Borner}},\ }\href {\doibase
  10.1088/0004-637x/707/1/354} {\bibfield  {journal} {\bibinfo  {journal}
  {Astrophysical Journal}\ }\textbf {\bibinfo {volume} {707}},\ \bibinfo
  {pages} {354} (\bibinfo {year} {2009})}\BibitemShut {NoStop}%
\bibitem [{\citenamefont {Mo}\ \emph {et~al.}(2010)\citenamefont {Mo},
  \citenamefont {van~den Bosch},\ and\ \citenamefont
  {White}}]{Mo:2010-Galaxy-formation-and-evolution}%
  \BibitemOpen
  \bibfield  {author} {\bibinfo {author} {\bibfnamefont {H.}~\bibnamefont
  {Mo}}, \bibinfo {author} {\bibfnamefont {F.}~\bibnamefont {van~den Bosch}}, \
  and\ \bibinfo {author} {\bibfnamefont {S.}~\bibnamefont {White}},\
  }\href@noop {} {\emph {\bibinfo {title} {Galaxy formation and evolution}}}\
  (\bibinfo  {publisher} {Cambridge University Press},\ \bibinfo {address}
  {Cambridge},\ \bibinfo {year} {2010})\BibitemShut {NoStop}%
\bibitem [{\citenamefont {Chan}\ \emph {et~al.}(2015)\citenamefont {Chan},
  \citenamefont {Kereš}, \citenamefont {Oñorbe}, \citenamefont {Hopkins},
  \citenamefont {Muratov}, \citenamefont {Faucher-Giguère},\ and\
  \citenamefont
  {Quataert}}]{Chan:2015-The-impact-of-baryonic-physics-on-the-structure}%
  \BibitemOpen
  \bibfield  {author} {\bibinfo {author} {\bibfnamefont {T.~K.}\ \bibnamefont
  {Chan}}, \bibinfo {author} {\bibfnamefont {D.}~\bibnamefont {Kereš}},
  \bibinfo {author} {\bibfnamefont {J.}~\bibnamefont {Oñorbe}}, \bibinfo
  {author} {\bibfnamefont {P.~F.}\ \bibnamefont {Hopkins}}, \bibinfo {author}
  {\bibfnamefont {A.~L.}\ \bibnamefont {Muratov}}, \bibinfo {author}
  {\bibfnamefont {C.-A.}\ \bibnamefont {Faucher-Giguère}}, \ and\ \bibinfo
  {author} {\bibfnamefont {E.}~\bibnamefont {Quataert}},\ }\href {\doibase
  10.1093/mnras/stv2165} {\bibfield  {journal} {\bibinfo  {journal} {Monthly
  Notices of the Royal Astronomical Society}\ }\textbf {\bibinfo {volume}
  {454}},\ \bibinfo {pages} {2981} (\bibinfo {year} {2015})},\ \Eprint
  {http://arxiv.org/abs/https://academic.oup.com/mnras/article-pdf/454/3/2981/4038253/stv2165.pdf}
  {https://academic.oup.com/mnras/article-pdf/454/3/2981/4038253/stv2165.pdf}
  \BibitemShut {NoStop}%
\bibitem [{\citenamefont {{Lelli}}\ \emph {et~al.}(2016)\citenamefont
  {{Lelli}}, \citenamefont {{McGaugh}},\ and\ \citenamefont
  {{Schombert}}}]{Lelli:2016-SPARC-Mass-Models-for-175-Disk-Galaxies}%
  \BibitemOpen
  \bibfield  {author} {\bibinfo {author} {\bibfnamefont {F.}~\bibnamefont
  {{Lelli}}}, \bibinfo {author} {\bibfnamefont {S.~S.}\ \bibnamefont
  {{McGaugh}}}, \ and\ \bibinfo {author} {\bibfnamefont {J.~M.}\ \bibnamefont
  {{Schombert}}},\ }\href {\doibase 10.3847/0004-6256/152/6/157} {\bibfield
  {journal} {\bibinfo  {journal} {\aj}\ }\textbf {\bibinfo {volume} {152}},\
  \bibinfo {eid} {157} (\bibinfo {year} {2016})},\ \Eprint
  {http://arxiv.org/abs/1606.09251} {arXiv:1606.09251 [astro-ph.GA]}
  \BibitemShut {NoStop}%
\bibitem [{\citenamefont {{Li}}\ \emph {et~al.}(2020)\citenamefont {{Li}},
  \citenamefont {{Lelli}}, \citenamefont {{McGaugh}},\ and\ \citenamefont
  {{Schombert}}}]{Li:2020-A-Comprehensive-Catalog-of-Dark-Matter-Halo-Models}%
  \BibitemOpen
  \bibfield  {author} {\bibinfo {author} {\bibfnamefont {P.}~\bibnamefont
  {{Li}}}, \bibinfo {author} {\bibfnamefont {F.}~\bibnamefont {{Lelli}}},
  \bibinfo {author} {\bibfnamefont {S.}~\bibnamefont {{McGaugh}}}, \ and\
  \bibinfo {author} {\bibfnamefont {J.}~\bibnamefont {{Schombert}}},\ }\href
  {\doibase 10.3847/1538-4365/ab700e} {\bibfield  {journal} {\bibinfo
  {journal} {\apjs}\ }\textbf {\bibinfo {volume} {247}},\ \bibinfo {eid} {31}
  (\bibinfo {year} {2020})},\ \Eprint {http://arxiv.org/abs/2001.10538}
  {arXiv:2001.10538 [astro-ph.GA]} \BibitemShut {NoStop}%
\bibitem [{\citenamefont {{Martinsson}}\ \emph {et~al.}(2013)\citenamefont
  {{Martinsson}}, \citenamefont {{Verheijen}}, \citenamefont {{Westfall}},
  \citenamefont {{Bershady}}, \citenamefont {{Andersen}},\ and\ \citenamefont
  {{Swaters}}}]{Martinsson:2014-The-DiskMass-Survey}%
  \BibitemOpen
  \bibfield  {author} {\bibinfo {author} {\bibfnamefont {T.~P.~K.}\
  \bibnamefont {{Martinsson}}}, \bibinfo {author} {\bibfnamefont {M.~A.~W.}\
  \bibnamefont {{Verheijen}}}, \bibinfo {author} {\bibfnamefont {K.~B.}\
  \bibnamefont {{Westfall}}}, \bibinfo {author} {\bibfnamefont {M.~A.}\
  \bibnamefont {{Bershady}}}, \bibinfo {author} {\bibfnamefont {D.~R.}\
  \bibnamefont {{Andersen}}}, \ and\ \bibinfo {author} {\bibfnamefont {R.~A.}\
  \bibnamefont {{Swaters}}},\ }\href {\doibase 10.1051/0004-6361/201321390}
  {\bibfield  {journal} {\bibinfo  {journal} {\aap}\ }\textbf {\bibinfo
  {volume} {557}},\ \bibinfo {eid} {A131} (\bibinfo {year} {2013})},\ \Eprint
  {http://arxiv.org/abs/1308.0336} {arXiv:1308.0336 [astro-ph.CO]} \BibitemShut
  {NoStop}%
\bibitem [{\citenamefont
  {Sofue}(2016)}]{Sofue:2016-Rotation-curve-decomposition-f}%
  \BibitemOpen
  \bibfield  {author} {\bibinfo {author} {\bibfnamefont {Y.}~\bibnamefont
  {Sofue}},\ }\href {\doibase 10.1093/pasj/psv103} {\bibfield  {journal}
  {\bibinfo  {journal} {Publications of the Astronomical Society of Japan}\
  }\textbf {\bibinfo {volume} {68}} (\bibinfo {year} {2016}),\
  10.1093/pasj/psv103}\BibitemShut {NoStop}%
\bibitem [{\citenamefont {Wang}\ and\ \citenamefont
  {Chen}(2021)}]{Wang:2021-Comparison-of-Modeling-SPARC-spiral-galaxies}%
  \BibitemOpen
  \bibfield  {author} {\bibinfo {author} {\bibfnamefont {L.}~\bibnamefont
  {Wang}}\ and\ \bibinfo {author} {\bibfnamefont {D.-M.}\ \bibnamefont
  {Chen}},\ }\href {\doibase 10.1088/1674-4527/21/11/271} {\bibfield  {journal}
  {\bibinfo  {journal} {Research in Astronomy and Astrophysics}\ }\textbf
  {\bibinfo {volume} {21}},\ \bibinfo {pages} {271} (\bibinfo {year}
  {2021})}\BibitemShut {NoStop}%
\bibitem [{\citenamefont {Kormendy}\ and\ \citenamefont
  {Freeman}(2004)}]{Kormendy:2004-Dark-Matter-in-Galaxies}%
  \BibitemOpen
  \bibfield  {author} {\bibinfo {author} {\bibfnamefont {J.}~\bibnamefont
  {Kormendy}}\ and\ \bibinfo {author} {\bibfnamefont {K.~C.}\ \bibnamefont
  {Freeman}},\ }in\ \href@noop {} {\emph {\bibinfo {booktitle} {Proc. IAU Symp.
  220, Dark Matter in Galaxies. Astron. Soc. Pac., San Francisco}}},\ \bibinfo
  {editor} {edited by\ \bibinfo {editor} {\bibfnamefont {W.~M.~A.}\
  \bibnamefont {Ryder S.~D.}, \bibfnamefont {Pisano D.~J.}}}\ (\bibinfo {year}
  {2004})\ p.\ \bibinfo {pages} {377}\BibitemShut {NoStop}%
\bibitem [{\citenamefont {Spano}\ \emph {et~al.}(2008)\citenamefont {Spano},
  \citenamefont {Marcelin}, \citenamefont {Amram}, \citenamefont {Carignan},
  \citenamefont {Epinat},\ and\ \citenamefont
  {Hernandez}}]{Spano:2008-GHASP-an-kinematic-survey-spiral}%
  \BibitemOpen
  \bibfield  {author} {\bibinfo {author} {\bibfnamefont {M.}~\bibnamefont
  {Spano}}, \bibinfo {author} {\bibfnamefont {M.}~\bibnamefont {Marcelin}},
  \bibinfo {author} {\bibfnamefont {P.}~\bibnamefont {Amram}}, \bibinfo
  {author} {\bibfnamefont {C.}~\bibnamefont {Carignan}}, \bibinfo {author}
  {\bibfnamefont {B.}~\bibnamefont {Epinat}}, \ and\ \bibinfo {author}
  {\bibfnamefont {O.}~\bibnamefont {Hernandez}},\ }\href {\doibase
  10.1111/j.1365-2966.2007.12545.x} {\bibfield  {journal} {\bibinfo  {journal}
  {Monthly Notices of the Royal Astronomical Society}\ }\textbf {\bibinfo
  {volume} {383}},\ \bibinfo {pages} {297} (\bibinfo {year} {2008})},\ \Eprint
  {http://arxiv.org/abs/https://academic.oup.com/mnras/article-pdf/383/1/297/3693757/mnras0383-0297.pdf}
  {https://academic.oup.com/mnras/article-pdf/383/1/297/3693757/mnras0383-0297.pdf}
  \BibitemShut {NoStop}%
\bibitem [{\citenamefont {{Barnes}}\ \emph {et~al.}(2004)\citenamefont
  {{Barnes}}, \citenamefont {{Sellwood}},\ and\ \citenamefont
  {{Kosowsky}}}]{Barnes:2004-Mass-Models-for-Spiral-Galaxies}%
  \BibitemOpen
  \bibfield  {author} {\bibinfo {author} {\bibfnamefont {E.~I.}\ \bibnamefont
  {{Barnes}}}, \bibinfo {author} {\bibfnamefont {J.~A.}\ \bibnamefont
  {{Sellwood}}}, \ and\ \bibinfo {author} {\bibfnamefont {A.}~\bibnamefont
  {{Kosowsky}}},\ }\href {\doibase 10.1086/425877} {\bibfield  {journal}
  {\bibinfo  {journal} {\aj}\ }\textbf {\bibinfo {volume} {128}},\ \bibinfo
  {pages} {2724} (\bibinfo {year} {2004})},\ \Eprint
  {http://arxiv.org/abs/astro-ph/0409239} {arXiv:astro-ph/0409239 [astro-ph]}
  \BibitemShut {NoStop}%
\bibitem [{\citenamefont {Chan}(2014)}]{Chan:2014-A-tight-scaling-relation}%
  \BibitemOpen
  \bibfield  {author} {\bibinfo {author} {\bibfnamefont {M.~H.}\ \bibnamefont
  {Chan}},\ }\href {\doibase 10.1093/mnrasl/slu047} {\bibfield  {journal}
  {\bibinfo  {journal} {Monthly Notices of the Royal Astronomical Society:
  Letters}\ }\textbf {\bibinfo {volume} {442}},\ \bibinfo {pages} {L14}
  (\bibinfo {year} {2014})},\ \Eprint
  {http://arxiv.org/abs/https://academic.oup.com/mnrasl/article-pdf/442/1/L14/9371164/slu047.pdf}
  {https://academic.oup.com/mnrasl/article-pdf/442/1/L14/9371164/slu047.pdf}
  \BibitemShut {NoStop}%
\bibitem [{\citenamefont
  {Xu}(2022{\natexlab{e}})}]{Xu:2022-The-origin-of-MOND-acceleratio}%
  \BibitemOpen
  \bibfield  {author} {\bibinfo {author} {\bibfnamefont {Z.}~\bibnamefont
  {Xu}},\ }\href {\doibase 10.48550/ARXIV.2203.05606} {\bibfield  {journal}
  {\bibinfo  {journal} {arXiv e-prints}\ ,\ \bibinfo {pages}
  {arXiv:2203.05606}} (\bibinfo {year} {2022}{\natexlab{e}})}\BibitemShut
  {NoStop}%
\bibitem [{\citenamefont {Markevitch}\ \emph {et~al.}(2004)\citenamefont
  {Markevitch}, \citenamefont {Gonzalez}, \citenamefont {Clowe}, \citenamefont
  {Vikhlinin}, \citenamefont {Forman}, \citenamefont {Jones}, \citenamefont
  {Murray},\ and\ \citenamefont
  {Tucker}}]{Markevitch:2004-Direct-Constraints-on-the-Dark-Matter}%
  \BibitemOpen
  \bibfield  {author} {\bibinfo {author} {\bibfnamefont {M.}~\bibnamefont
  {Markevitch}}, \bibinfo {author} {\bibfnamefont {A.~H.}\ \bibnamefont
  {Gonzalez}}, \bibinfo {author} {\bibfnamefont {D.}~\bibnamefont {Clowe}},
  \bibinfo {author} {\bibfnamefont {A.}~\bibnamefont {Vikhlinin}}, \bibinfo
  {author} {\bibfnamefont {W.}~\bibnamefont {Forman}}, \bibinfo {author}
  {\bibfnamefont {C.}~\bibnamefont {Jones}}, \bibinfo {author} {\bibfnamefont
  {S.}~\bibnamefont {Murray}}, \ and\ \bibinfo {author} {\bibfnamefont
  {W.}~\bibnamefont {Tucker}},\ }\href {\doibase 10.1086/383178} {\bibfield
  {journal} {\bibinfo  {journal} {The Astrophysical Journal}\ }\textbf
  {\bibinfo {volume} {606}},\ \bibinfo {pages} {819} (\bibinfo {year}
  {2004})}\BibitemShut {NoStop}%
\bibitem [{\citenamefont {{Brada{\v{c}}}}\ \emph {et~al.}(2008)\citenamefont
  {{Brada{\v{c}}}}, \citenamefont {{Allen}}, \citenamefont {{Treu}},
  \citenamefont {{Ebeling}}, \citenamefont {{Massey}}, \citenamefont
  {{Morris}}, \citenamefont {{von der Linden}},\ and\ \citenamefont
  {{Applegate}}}]{Brada:2008-Revealing-the-Properties-of-Dark-Matter}%
  \BibitemOpen
  \bibfield  {author} {\bibinfo {author} {\bibfnamefont {M.}~\bibnamefont
  {{Brada{\v{c}}}}}, \bibinfo {author} {\bibfnamefont {S.~W.}\ \bibnamefont
  {{Allen}}}, \bibinfo {author} {\bibfnamefont {T.}~\bibnamefont {{Treu}}},
  \bibinfo {author} {\bibfnamefont {H.}~\bibnamefont {{Ebeling}}}, \bibinfo
  {author} {\bibfnamefont {R.}~\bibnamefont {{Massey}}}, \bibinfo {author}
  {\bibfnamefont {R.~G.}\ \bibnamefont {{Morris}}}, \bibinfo {author}
  {\bibfnamefont {A.}~\bibnamefont {{von der Linden}}}, \ and\ \bibinfo
  {author} {\bibfnamefont {D.}~\bibnamefont {{Applegate}}},\ }\href {\doibase
  10.1086/591246} {\bibfield  {journal} {\bibinfo  {journal} {\apj}\ }\textbf
  {\bibinfo {volume} {687}},\ \bibinfo {pages} {959} (\bibinfo {year}
  {2008})},\ \Eprint {http://arxiv.org/abs/0806.2320} {arXiv:0806.2320
  [astro-ph]} \BibitemShut {NoStop}%
\bibitem [{\citenamefont {{Dawson}}\ \emph {et~al.}(2012)\citenamefont
  {{Dawson}}, \citenamefont {{Wittman}}, \citenamefont {{Jee}}, \citenamefont
  {{Gee}}, \citenamefont {{Hughes}}, \citenamefont {{Tyson}}, \citenamefont
  {{Schmidt}}, \citenamefont {{Thorman}}, \citenamefont {{Brada{\v{c}}}},
  \citenamefont {{Miyazaki}}, \citenamefont {{Lemaux}}, \citenamefont
  {{Utsumi}},\ and\ \citenamefont
  {{Margoniner}}}]{Dawson:2012-Discovery-of-a-Dissociative-Galaxy}%
  \BibitemOpen
  \bibfield  {author} {\bibinfo {author} {\bibfnamefont {W.~A.}\ \bibnamefont
  {{Dawson}}}, \bibinfo {author} {\bibfnamefont {D.}~\bibnamefont {{Wittman}}},
  \bibinfo {author} {\bibfnamefont {M.~J.}\ \bibnamefont {{Jee}}}, \bibinfo
  {author} {\bibfnamefont {P.}~\bibnamefont {{Gee}}}, \bibinfo {author}
  {\bibfnamefont {J.~P.}\ \bibnamefont {{Hughes}}}, \bibinfo {author}
  {\bibfnamefont {J.~A.}\ \bibnamefont {{Tyson}}}, \bibinfo {author}
  {\bibfnamefont {S.}~\bibnamefont {{Schmidt}}}, \bibinfo {author}
  {\bibfnamefont {P.}~\bibnamefont {{Thorman}}}, \bibinfo {author}
  {\bibfnamefont {M.}~\bibnamefont {{Brada{\v{c}}}}}, \bibinfo {author}
  {\bibfnamefont {S.}~\bibnamefont {{Miyazaki}}}, \bibinfo {author}
  {\bibfnamefont {B.}~\bibnamefont {{Lemaux}}}, \bibinfo {author}
  {\bibfnamefont {Y.}~\bibnamefont {{Utsumi}}}, \ and\ \bibinfo {author}
  {\bibfnamefont {V.~E.}\ \bibnamefont {{Margoniner}}},\ }\href {\doibase
  10.1088/2041-8205/747/2/L42} {\bibfield  {journal} {\bibinfo  {journal}
  {\apjl}\ }\textbf {\bibinfo {volume} {747}},\ \bibinfo {eid} {L42} (\bibinfo
  {year} {2012})},\ \Eprint {http://arxiv.org/abs/1110.4391} {arXiv:1110.4391
  [astro-ph.CO]} \BibitemShut {NoStop}%
\bibitem [{\citenamefont {Robertson}\ \emph {et~al.}(2016)\citenamefont
  {Robertson}, \citenamefont {Massey},\ and\ \citenamefont
  {Eke}}]{Robertson:2016-What-does-the-Bullet-Cluster-tell}%
  \BibitemOpen
  \bibfield  {author} {\bibinfo {author} {\bibfnamefont {A.}~\bibnamefont
  {Robertson}}, \bibinfo {author} {\bibfnamefont {R.}~\bibnamefont {Massey}}, \
  and\ \bibinfo {author} {\bibfnamefont {V.}~\bibnamefont {Eke}},\ }\href
  {\doibase 10.1093/mnras/stw2670} {\bibfield  {journal} {\bibinfo  {journal}
  {Monthly Notices of the Royal Astronomical Society}\ }\textbf {\bibinfo
  {volume} {465}},\ \bibinfo {pages} {569} (\bibinfo {year} {2016})},\ \Eprint
  {http://arxiv.org/abs/https://academic.oup.com/mnras/article-pdf/465/1/569/8593057/stw2670.pdf}
  {https://academic.oup.com/mnras/article-pdf/465/1/569/8593057/stw2670.pdf}
  \BibitemShut {NoStop}%
\bibitem [{\citenamefont
  {Kolmogoroff}(1941)}]{Kolmogoroff:1941-Dissipation-of-energy-in-the-l}%
  \BibitemOpen
  \bibfield  {author} {\bibinfo {author} {\bibfnamefont {A.~N.}\ \bibnamefont
  {Kolmogoroff}},\ }\href {<Go to ISI>://WOS:000201918500005} {\bibfield
  {journal} {\bibinfo  {journal} {Comptes Rendus De L Academie Des Sciences De
  L Urss}\ }\textbf {\bibinfo {volume} {32}},\ \bibinfo {pages} {16} (\bibinfo
  {year} {1941})}\BibitemShut {NoStop}%
\bibitem [{\citenamefont
  {Xu}(2022{\natexlab{f}})}]{Xu:2022-Dark_matter-flow-dataset-part1}%
  \BibitemOpen
  \bibfield  {author} {\bibinfo {author} {\bibfnamefont {Z.}~\bibnamefont
  {Xu}},\ }\href {\doibase 10.5281/zenodo.6541230} {\enquote {\bibinfo {title}
  {Dark matter flow dataset part i: Halo-based statistics from cosmological
  n-body simulation},}\ } (\bibinfo {year} {2022}{\natexlab{f}})\BibitemShut
  {NoStop}%
\bibitem [{\citenamefont
  {Xu}(2022{\natexlab{g}})}]{Xu:2022-Dark_matter-flow-dataset-part2}%
  \BibitemOpen
  \bibfield  {author} {\bibinfo {author} {\bibfnamefont {Z.}~\bibnamefont
  {Xu}},\ }\href {\doibase 10.5281/zenodo.6569898} {\enquote {\bibinfo {title}
  {Dark matter flow dataset part ii: Correlation-based statistics from
  cosmological n-body simulation},}\ } (\bibinfo {year}
  {2022}{\natexlab{g}})\BibitemShut {NoStop}%
\end{thebibliography}%

\label{lastpage}
\end{document}